\documentclass{aa}
\usepackage{txfonts}
\usepackage{graphicx}
\usepackage{rotating}
\usepackage{natbib}
\usepackage{lscape}
\usepackage{multirow}

\bibpunct{(}{)}{;}{a}{}{,} %to follow the A&A style

\newcommand{\gapprox}{$\stackrel {>}{_{\sim}}$}   %greater/less approx.
\newcommand{\lapprox}{$\stackrel {<}{_{\sim}}$}
\newcommand{\av}{$A_\mathrm{V}$}
\newcommand{\kms}{km\,s$^{-1}$}

\newcommand{\lsun}{$L_{\odot}$}

\newcommand{\Te}{$T_\mathrm{e}$}
\newcommand{\dens}{$n_\mathrm{e}$}
\newcommand{\xe}{$x_\mathrm{e}$}
\newcommand{\nH}{$n_\mathrm{H}$}

\newcommand{\oi}{\ion{O}{i}}

\newcommand{\si}{\ion{S}{i}}
\newcommand{\sii}{\ion{S}{ii}}
\newcommand{\siii}{\ion{S}{iii}}
\newcommand{\n}{\ion{N}{i}}
\newcommand{\nii}{\ion{N}{ii}}

\newcommand{\feii}{\ion{Fe}{ii}}

\newcommand{\pii}{\ion{P}{ii}}
\newcommand{\li}{\ion{Li}{i}}
\begin{document}

\title{GIARPS High-resolution Observations of  T Tauri stars (GHOsT).\\ I: Jet line emission }

\author{T. Giannini\inst{1},
B. Nisini\inst{1},
S. Antoniucci\inst{1},
K. Biazzo\inst{1,2},
J. Alcal{\'a}\inst{3},
F. Bacciotti\inst{4},   
D. Fedele\inst{4},            
A. Frasca\inst{2},
A. Harutyunyan\inst{5},
U. Munari\inst{6},
E. Rigliaco\inst{6},
F. Vitali\inst{1}
}
\institute{ 
INAF - Osservatorio Astronomico di Roma - Via Frascati 33, 00078, Monte Porzio Catone, Italy
\and
INAF - Osservatorio Astrofisico di Catania -  Via S.Sofia 78, 95123, Catania, Italy 
\and
INAF - Osservatorio Astronomico di Capodimonte - Salita Moiariello 16, 80131, Napoli, Italy
\and                    
INAF - Osservatorio Astrofisico di Arcetri - Largo Enrico Fermi 5, 50125, Firenze, Italy
\and
Fundaci\'{o}n Galileo Galilei – INAF - Telescopio Nazionale Galileo, 38700, Bre\~{n}a Baja, Santa Cruz de Tenerife, Spain
\and
INAF - Osservatorio Astronomico di Padova -  Via dell' Osservatorio 8, 36012, Asiago (VI), Italy
}
\offprints{Teresa Giannini, \email{teresa.giannini@inaf.it}}
\date{Received date / Accepted date}
\titlerunning{GHOsT project: jet emission}
\authorrunning{Giannini et al.}

\abstract
{The mechanism for jet formation in the disks of T Tauri stars remains poorly understood. Observational benchmarks to launching models can be provided by tracing the physical properties of the  kinematic components of the wind and jet in the inner 100 au of the disk surface.}
{In the framework of the GHOsT (GIARPS High-resolution Observations of T Tauri stars) project, we aim to perform a multi-line analysis of the velocity components of the gas in the jet acceleration zone.}
{We analyzed the GIARPS-TNG spectra of six objects in the Taurus-Auriga complex (RY Tau, DG Tau, DL Tau, HN Tau, DO Tau, RW Aur A). Thanks to the combined high-spectral resolution
($\Re$\,=\.50\,000$-$115\,000) and wide spectral coverage ($\sim$\,400$-$2400 nm) we observed several {\ion{O}{$^0$}}, {\ion{S}{$^+$}}, {\ion{N}{$^0$}}, {\ion{N}{$^+$}}, and {\ion{Fe}{$^+$}}  forbidden lines spanning a large range of excitation and ionization conditions. In four objects (DG Tau, HN Tau, DO Tau, RW Aur A), temperature (\Te), electron and total density (\dens,\,\nH), and fractional ionization (\xe) were derived as a function of velocity through an excitation and ionization model. The abundance of gaseous iron, X(Fe), a probe of the dust content in the jet, was derived in selected velocity channels.} 
{The physical parameters vary smoothly with velocity, suggesting a common origin for the different kinematic components. In DG Tau and HN Tau, \Te,\, \xe\,, and X(Fe)
 increase with velocity (roughly from 6000 K, 0.05, 10\%X(Fe)$_\odot$ to  15\,000 K, 0.6, 90\%X(Fe)$_\odot$). This trend is in agreement with disk--wind models in
  which the jet is launched from regions of the disk at different radii. In DO Tau and RW Aur A, we infer \xe $<$ 0.1, \nH $\sim$ 10$^{6-7}$ cm$^{-3}$,  and X(Fe)\,\lapprox\,X(Fe)$_\odot$ at all velocities. These findings are tentatively explained by the formation of these jets from dense regions inside the inner, gaseous disk, or as a consequence of their high degree of collimation.}
{}
\keywords{Line: profiles -- Stars: pre-main sequence -- Stars: jets -- Stars:individual (RY Tau, DG Tau, DL Tau, HN Tau, DO Tau, RW Aur A)}

\maketitle

\section{Introduction}\label{sec:sec1}
The formation of stars through disk accretion is often accompanied by mass outflows in the form of high-velocity collimated jets. The tight relationship between mass accretion and mass outflow in the jets has been so far demonstrated in populations of low-mass pre-main sequence stars
(Classical T Tauri, or CTT; e.g., Nisini et al. 2018, and references therein). In spite of that, the exact mechanism driving the jets and the region of the disk involved in such a process remains poorly understood,  although there is a consensus that a magneto-hydrodynamic (MHD) mechanism should be at the base of any model (e.g., Frank et al. 2014). In addition to collimated high-velocity jets, matter  also escapes from the disk surface in the form of slow uncollimated winds, driven either by the same MHD mechanism causing jets or by photo-evaporation due to  UV/X-ray photons from the star (e.g., Ercolano and Pascucci, 2017). Such winds play a fundamental role in disk dispersal during the late stage of CTT evolution.
Observationally, optical forbidden lines, especially the [\oi]630 nm line\footnote{Hereinafter all the lines are identified with their wavelength expressed in nm.}, have been widely used to study both jets and winds from CTT stars.
High-resolution spectroscopic observations have shown that these lines can be deconvolved into separate kinematic components tracing different wind structures. In particular, a low-velocity component (LVC, |$v$| \lapprox\, 40\, km\,s$^{-1}$), is attributed to slow uncollimated winds, while a high-velocity component (HVC, |$v$| \gapprox\, 40 km\,s$^{-1}$) represents the collimated jet (e.g., Hartigan et al. 1995). The LVC is sometimes observed not as a single Gaussian profile but as the sum of two (or more) components attributed to winds originating in different disk regions (Rigliaco et al. 2013, Simon et al. 2016).
Several studies have addressed the properties of these different line emission components in an attempt to constrain their
origin and their role in the  evolution of  disk dispersal. 
Optical high-resolution observations of  CTT stars have mainly addressed the kinematic properties of the [\oi]630 line or performed a diagnostic analysis based on very few lines  and exploiting a limited diagnostic range
(e.g., Fang et al. 2018, Banzatti et al. 2019). 
On the other hand, recent VLT/X-shooter observations have shown the potential of combining several diagnostic forbidden lines in the optical and infrared (IR) ranges to  constrain  the physical properties of the emitting gas associated with the jets (Bacciotti et al. 2011, Giannini et al. 2013, Nisini et al. 2016). 

Ideally, a more detailed understanding of the origin of the various mass-loss phenomena in disks would come from the coupling of high spectral resolution and wide wavelength coverage. The GIARPS (GIAno and haRPS; Claudi et al. 2017) observing mode at the {\it Telescopio Nazionale Galileo} (TNG) can now achieve this task by combining HARPS-N and GIANO-B visible and near-infrared (NIR) high-resolution spectrographs, offering a unique combination of wide spectral coverage (390$-$690 nm plus 900$-$2420 nm) and high spectral resolution ($\Re\sim$\,115\,000 and 50\,000). We can therefore take advantage of the GIARPS capabilities to infer detailed excitation conditions of gas associated with kinematically separated wind and jet components. In particular, the simultaneous use of several optical and IR forbidden lines allows one to determine important gas physical quantities, such as temperature, density, and ionization degree.  The best tracers are strong lines of the most abundant atomic species (such as O, S, N, Fe), usually neutral or in the first ionization stage. The intensities and intensity ratios of these lines are used to probe the excitation conditions and chemical composition inside the jet beam  (e.g., Giannini et al. 2013, 2015, Podio et al. 2009, Hartigan et al. 2004, Bacciotti and Eisl{\"o}ffel 1999). Also, in a few cases, the variation of the physical quantities with the gas velocity has been analyzed (Maurri et al. 2014, Garcia Lopez et al. 2008,  Coffey et al. 2008).  This may place fundamental observational constraints on jet-launching models and may give indications regarding the part of the protostellar-disk system from which winds and jets originate.

Another fundamental issue to address is the dust content in the jet beam. Both observations and theoretical models show
that in the unperturbed medium several refractory species, like iron, titanium, magnesium, silicon, and calcium,  are depleted into dust grains, and  their gas-phase abundance is  up to a factor 10$^2-$10$^4$ lower with respect to the solar one (e.g., Savage and Sembach 1996, May et al. 2000, Mouri and Taniguchi 2000). 
Gas--grain and grain--grain collisions occurring in shock waves may erode and/or vaporize the dust (e.g., Jones 2000), although
total dust disruption is expected only in high-velocity shocks ($v_\textrm{shock}$ $>$ 200 km\,s$^{-1}$), such as those produced by supernovae explosions. In protostellar jets, the typical shock velocities are small ($v_\textrm{shock} \sim$ 20$-$50 km\,s$^{-1}$), and therefore only a partial disruption of the dust is expected. Moreover, the depletion degree may also depend on the distance of the jet launching region from the star. If this region is in the proximity of the inner edge of the disk,  at distances from the star of less than the dust sublimation radius (typically 0.1 au, X-winds, Shu et al. 2000), the abundance of the refractory species is expected to be similar to the solar value. Sub-solar values are found if the jet comes from the external, dusty parts of the disk (disk-winds, K{\"o}nigl et al. 2010). Therefore, the measurement of the depletion degree in the different velocity components is crucial to shed light on jet formation, as it depends upon both the location of the launching region and the ability of the jet to destroy the dusty disk material.
Several observations have indicated that a significant degree of dust depletion occurs in the jet region close to the disk (e.g., Nisini et al. 2005, Podio et al. 2006, Giannini et al. 2013), and that the depletion might decrease with jet velocity (Agra-Amboage et al. 2011, Nisini et al. 2016), suggesting that high-velocity jet channels preferentially originate in dust-free disk regions.

In the present paper, we investigate the two aspects described above in a sample of six CTTs known to drive atomic jets. These have been observed with GIARPS as part of the GHOsT (GIARPS High-resolution Observations of T Tauri stars)  project, which has the broader aim of characterizing the star--disk interaction region of CTTs in the Taurus-Auriga star-forming region. The first results showing the potential of high-resolution IR observations obtained with the GIANO-B instrument have been presented in Antoniucci et al. (2017).

Considering the distance of Taurus-Auriga (between 120 and 160 pc) and the spatial resolution of GIARPS, we can probe the gas at a distance of about 100 au from the exciting source, where the collimation of the jet takes place. Also, this distance is sufficiently small to assume that the outflowing gas, not yet having interacted enough with the environmental medium, still preserves the initial physical conditions of the launching region.  These observations represent a testbed for studies of a larger sample to be performed in the coming years. Our first goal is to assess the GIARPS capability in detecting jet lines. These are then analyzed to 
obtain a detailed picture of the gas excitation and ionization conditions as a function of velocity, as well as to estimate the dust content in the different jet components.

The paper is organized as follows. In Sect. \ref{sec:sec2} we describe the observations and data reduction. Profiles of the observed lines are shown in Sect. \ref{sec:sec3}, while the data analysis is presented in Sect. \ref{sec:sec4}. A discussion of our results is given in Sect. \ref{sec:sec5}, while a summary of the main results is presented in Sect.  \ref{sec:sec6}. Finally, in Appendix A we describe our excitation/ionization model.

\section{Observations and data reduction}\label{sec:sec2}

\begin{table*}
\caption{\label{tab:tab1} Source properties.}
\footnotesize
\begin{tabular}{ccccccccc}
\hline\hline  
Source  & d                & SpT             & A$_V$                & $L_{\textrm{*}}$        & $L_{\mathrm{acc}}$  & $M_{\textrm{*}}$       & log($M_{\mathrm{acc}}$)     & $i_\textrm{disk}$    \\
\hline
        & (pc)             &                 & (mag)                & (L$_{\odot}$)           & (L$_{\odot}$)       & (M$_{\odot}$)          & (M$_{\odot}$\,yr$^{-1}$)    & (deg)         \\
\hline\hline  
RY Tau  & 128$\pm$4$^1$    & F7$^1$         & 1.8$^2$               &  10.7$^2$               &   -                 & 2.04$^3$               &  -7.3$^4$                   & 65$^1$         \\
DG Tau  & 121$\pm$18$^5$   & K7.0$^2$       & 1.60$^2$              &  0.5$^2$                & 0.09$^6$            & 0.76$^6$               &  -8.12$^6$                  & 37$^7$         \\
DL Tau  & 159$\pm$8$^5$    & K5.5$^2$       & 1.80$^2$              &  0.5$^2$                & 0.14$^6$            & 0.75$^8$               &  -8.06$^6$                  & 45$^1$         \\
HN Tau  & 145$\pm$19$^5$   & K3$^2$         & 1.15$^2$              &  0.16$^2$               & 0.05$^6$            & 0.78$^8$               &  -8.69$^6$                  & 70$^3$         \\
DO Tau  & 139$\pm$7$^5$    & M0.3$^2$       & 0.75$^2$              &  0.22$^2$               & 0.1$^6$             & 0.56$^8$               &  -8.21$^6$                  & 28$^3$         \\   
RW Aur A& 163$\pm$10$^5$   & K0$^2$         & 0$-$2$^{2,9}$         &  0.72$^2$               &   -                 & 1.48$^8$               &  -7.39$^{10}$               & 55$^3$         \\
 \hline  
\end{tabular}
\tablebib{(1) \citet{2018ApJ...869...17L}; (2)\,\citet{2014ApJ...786...97H}:  (3)\,\citet{2019ApJ...882...49L}; (4)\,\citet{2018ApJ...855..143S};
(5)\,\citet{2018A&A...616A...1G}; (6)\,\citet{2016ApJ...831..169S};
(7)\,\citet{2018ApJ...865L..12B}; (8) Rigliaco et al. 2015 and references therein; (9)\,\citet{2019A&A...625A..49K}; (10) \citet{2016A&A...596A..38F}.}
\end{table*}

\subsection{TNG/GIARPS observations and spectra extraction}
The observed sample consists of six sources selected among the brightest CTTs of the Taurus-Auriga star-forming region. These are known to drive a high-velocity jet, observed either by direct imaging or by the presence of a high-velocity component in their forbidden optical lines (e.g., Agra-Amboage et al. 2009, Coffey et al. 2012, Simon et al. 2016,  Skinner et al. 2018).
Table\,\ref{tab:tab1} lists some relevant stellar parameters taken from the recent literature: distance, spectral type, visual extinction, stellar and accretion luminosity, stellar mass, mass accretion rate, and disk inclination.

The targets were observed in 2017 on two nights (October 29 and November 13), using the GIARPS observing mode (\citealt{claudietal2017}), 
at the 3.58\,m {\it Telescopio Nazionale Galileo} (TNG; La Palma, Spain). The GIARPS mode is enabled by the optical coupling of the TNG HARPS-N and GIANO-B 
spectrographs installed at the telescope Nasmyth `B' focus through a dichroic splitting of the visible and NIR light. The average seeing was 0\farcs 8 on both nights.

HARPS-N is the visible high-resolution spectrograph mounted at the TNG, and covers the spectral domain from 390 to 690\,nm with a 
mean resolution of $\Re$=115\,000 (\citealt{cosentinoetal2012}). HARPS-N is equipped with two fibers (FoV\,=\,1$\arcsec$). We decided to place a fiber on the target and the other on the sky. Total exposure times are reported in Table~\ref{tab:tab2}. The reduction of the spectra was obtained using the latest version (Nov. 2013) 
of the HARPS-N instrument Data Reduction Software pipeline\footnote{See details in http://www.tng.iac.es/instruments/harps/.} and applying the appropriate mask depending on the spectral type of the target (e.g., \citealt{pepeetal2002}). The basic processing steps for the data reduction consist of the bias and dark current subtraction, 
flat fielding, wavelength calibration, spectrum extraction, and cross-correlation computation. We used an interactive procedure for the removal of the spurious features caused by the atmosphere of the Earth  (Frasca et al. 2000). We adopted as telluric templates HARPS-N spectra of hot, fast-rotating stars, where the broad and shallow photospheric absorption lines have been purposely flattened.
We then used the tool {\sc rotfit} (Frasca et al. 2003) to fit the photospheric absorption spectra of our targets with a grid of templates and derive in a self-consistent way the atmospheric parameters, the projected rotational velocity ($v\sin i$), and the line veiling resulting from the accretion shock on the stellar surface (Alcal\'a et al., in preparation). 
For each target, the rotationally broadened photospheric template was subtracted from the observed spectrum to clean the emission lines from nearby photospheric features. This is important to prepare the data for the following analysis, especially for the profile of the [\oi]630 line.

GIANO-B is the NIR high-resolution spectrograph of the TNG (\citealt{olivaetal2012, origliaetal2014}). 
The instrument provides cross dispersed echelle spectroscopy at a resolving power of $\Re$=50\,000 over the 900$-$2420\,nm spectral range in a single exposure. 
The telescope light is directly fed on the instrument slit, which has on-sky dimensions of $6\arcsec \times 0.5\arcsec$.
The spectra were acquired with the classical nodding-on-slit technique, that is, by alternately observing the target at two different positions (A and B) along the slit. 
The subtraction of these two consecutive exposures ensures optimal removal of both sky emission and instrumental background. Total exposure times are reported in Table~\ref{tab:tab2}. The spectra extraction was performed following the 2D GIANO-B data reduction 
prescriptions\footnote{More details can be found on the GIANO-B website http://www.tng.iac.es/instruments/giano/.}, as also described for example in \cite{carleoetal2018}. 
Halogen lamp exposures were employed to map the order geometry and for flat-field correction, while wavelength calibration was based on lines from a U-Ne lamp 
acquired at the end of each night.
For the removal of telluric lines in the NIR we used the tool {\sc molecfit}, which combines a radiative transfer code, a molecular line database, atmospheric profiles, and various kernels to model the instrument line spread function (\citealt{smetteetal2015}). 
In the end, a synthetic sky was modeled independently for each GIANO-B spectral order of our interest. We then used the IRAF\footnote{IRAF is distributed by the National Optical Astronomy Observatory, which is operated by the Association of the Universities for Research in Astronomy, inc., under cooperative agreement with the National Science Foundation.} task {\sc telluric} to correct the target spectra for the telluric absorptions. This procedure consists of shifting and scaling the synthetic telluric spectrum to the target spectrum to best divide out the telluric features from the target spectrum.

For each object, we measured the shift of the velocity scale with respect to the local standard of rest-frame in the HARPS-N spectra using the profile of the \li\, photospheric doublet, assuming weighted $\lambda_\textrm{air}$\,=\,670.7876 nm. The shift is between 15 km\,s$^{-1}$ (in DL Tau) and 22 km\,s$^{-1}$ (in RW Aur), and was applied also to the GIANO-B spectra as they were acquired simultaneously with the HARPS-N ones.

\subsection{Ancillary low-resolution spectroscopy and photometry}
To flux calibrate the HARPS-N spectra, we collected low-resolution ($\Re$=2400) optical spectra of our sources during the nights of 27 October and 11 November, 2017. The observations were obtained with the 1.22\,m telescope operated in Asiago (Italy) by the University of Padova.  The spectra cover the wavelength interval 330$-$790 nm. They were fully reduced and flux calibrated against a spectrophotometric standard, and their flux zero-point was checked against the $BVR_{\textrm C}I_{\textrm C}$ photometric measurements (Table\,\ref{tab:tab2}), collected on the same nights with the {\it Asiago Novae \& Symbiotic Stars Collaboration} (ANS) telescopes (Munari et al. 2012).

Near-infrared photometry of the objects in the $JHK_s$ bands was obtained with the REMIR camera on the {\it Rapid Eye Mount} (REM) telescope (Vitali et al. 2003), located at the La Silla Observatory (Chile), on the night of 11 November, 2017 (see Table~\ref{tab:tab2}). 

\subsection{Flux calibration of the GIARPS spectra}
HARPS-N spectra were flux calibrated using the Asiago low-resolution spectra, which were taken within two nights before our GHOsT run. Given the short temporal distance between the two data sets we assume that the continuum shape did not change significantly between the Asiago and TNG observations.
On this basis, for each source we first fitted the continuum of the Asiago spectrum and then multiplied it for the continuum-normalized HARPS-N spectrum.

For the flux calibration of GIANO-B spectra, we took into account the collected photometric points in the $I_CJHK_{\textrm s}$ bands, assuming that the magnitudes did not change significantly between TNG, Asiago, and REM observations. We then performed an interpolation between the considered flux measurements using a spline function to derive a smooth continuum function in the interval 940$-$2420 nm, which we employed to flux-calibrate the various (continuum-normalized) segments of the GIANO-B spectrum. 

\begin{table*}
\caption{\label{tab:tab2} Journal of observations and photometry.}
\centering
\begin{tabular}{ccccccccccc}
\hline\hline
Source     &   Obs  Date    &  $t_{\textrm exp}$-HARPS-N & $t_{\textrm exp}$-GIANO-B     &  $B$    &   $V$    & $R_\textrm{C}$ & $I_\textrm {C}$ &  $J$  &   $H$   &   $K_\textrm {s}$   \\
\hline
           &                &   (s)        &       (s)       &  (mag)  &  (mag)   &  (mag)      &  (mag)      &  (mag)&  (mag)  &  (mag)  \\
\hline\hline
RY Tau     &   13 Nov 2017  &   1500       &   1200           & 11.49   & 10.36    &  9.61       &   8.81      &  6.95 &   6.45  &   5.84  \\
DG Tau     &   29 Oct 2017  &   2200       &   1800           & 13.65   & 12.58    & 11.50       &  10.46      &  -    &   7.66  &   6.80  \\
DL Tau     &   29 Oct 2017  &   3000       &   2400           & 14.32   & 13.06    & 12.05       &  11.01      &  9.55 &   8.61  &   7.93  \\
HN Tau     &   29 Oct 2017  &   4500       &   3600           & 15.00   & 13.99    & 13.12       &  12.27      & 10.82 &   9.79  &   8.78  \\
DO Tau     &   13 Nov 2017  &   3000       &   2400           & 14.34   & 13.18    & 12.27       &  11.23      &  9.28 &   8.17  &   7.34  \\
RW Aur A   &   13 Nov 2017  &   2200       &   1800           & 11.04   & 10.44    &  9.97       &   9.38      &  8.41 &   7.66  &   7.06  \\
\hline
\end{tabular}
\tablefoot{Typical errors in photometric magnitudes are 0.01 mag in the optical bands and 0.02 mag in the NIR bands.}
\end{table*}

\section{Results}\label{sec:sec3}
\subsection{Line profiles: description and comparison with the literature}\label{sec:sec3.1}

\begin{figure*}
   \centering
   \includegraphics[width=10cm]{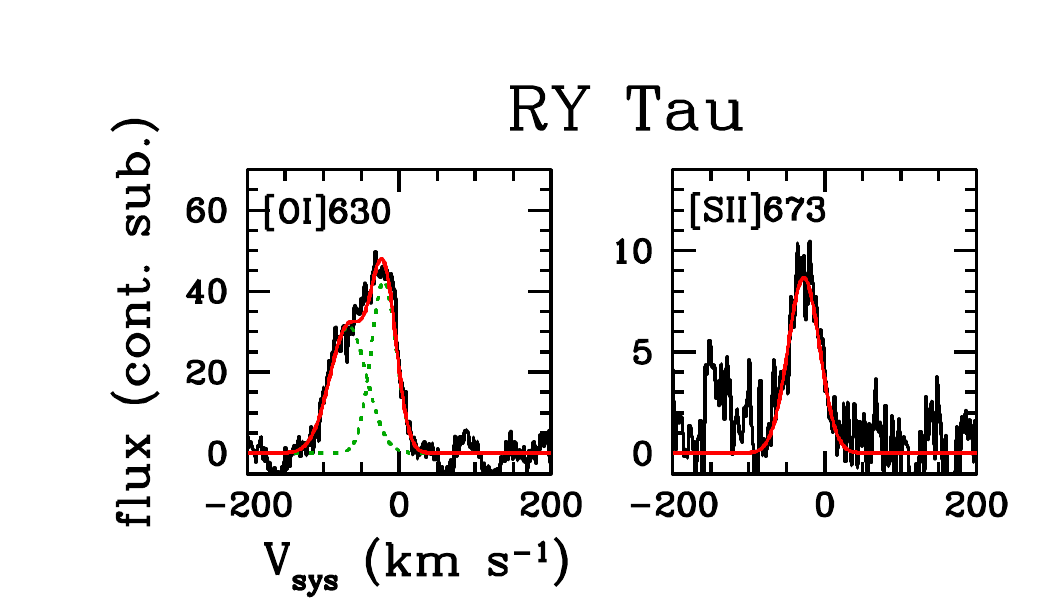}
   \caption{ \label{fig:fig1} Continuum-subtracted line profiles of the forbidden lines detected in RY Tau (black). In red is the fit to the profile, obtained by adding
   multiple Gaussians (green dotted lines). Flux units are 10$^{-15}$erg s$^{-1}$ cm$^{-2}$ \AA$^{-1}$, while the line wavelength reported in the label is in nanometers.}
 \end{figure*}

\begin{figure*}
   \centering
   \includegraphics[width=18cm]{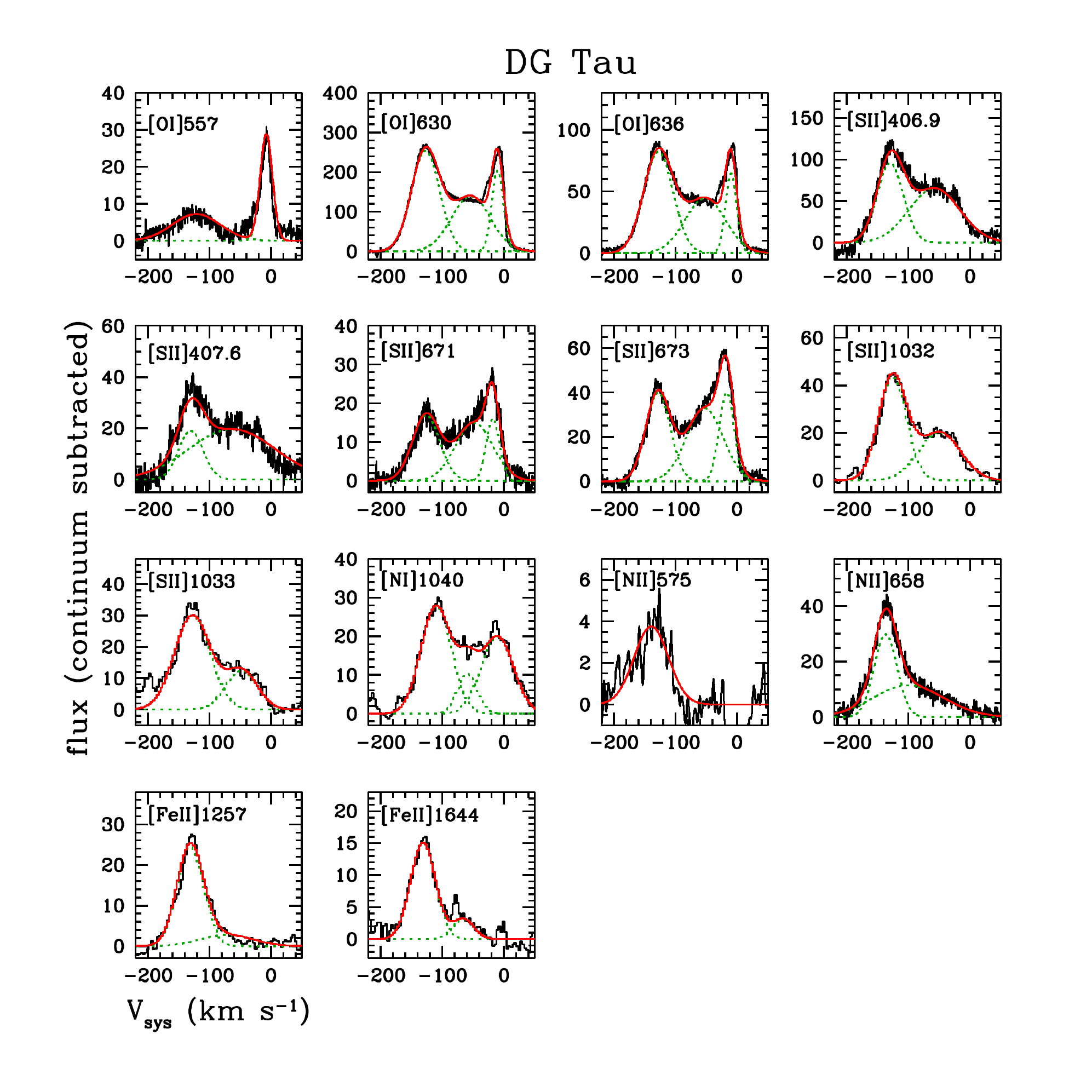}
   \caption{ \label{fig:fig2} As in Fig.\,\ref{fig:fig1} but for  DG Tau.}
 \end{figure*}

\begin{figure*}
  \centering
  \includegraphics[width=18cm]{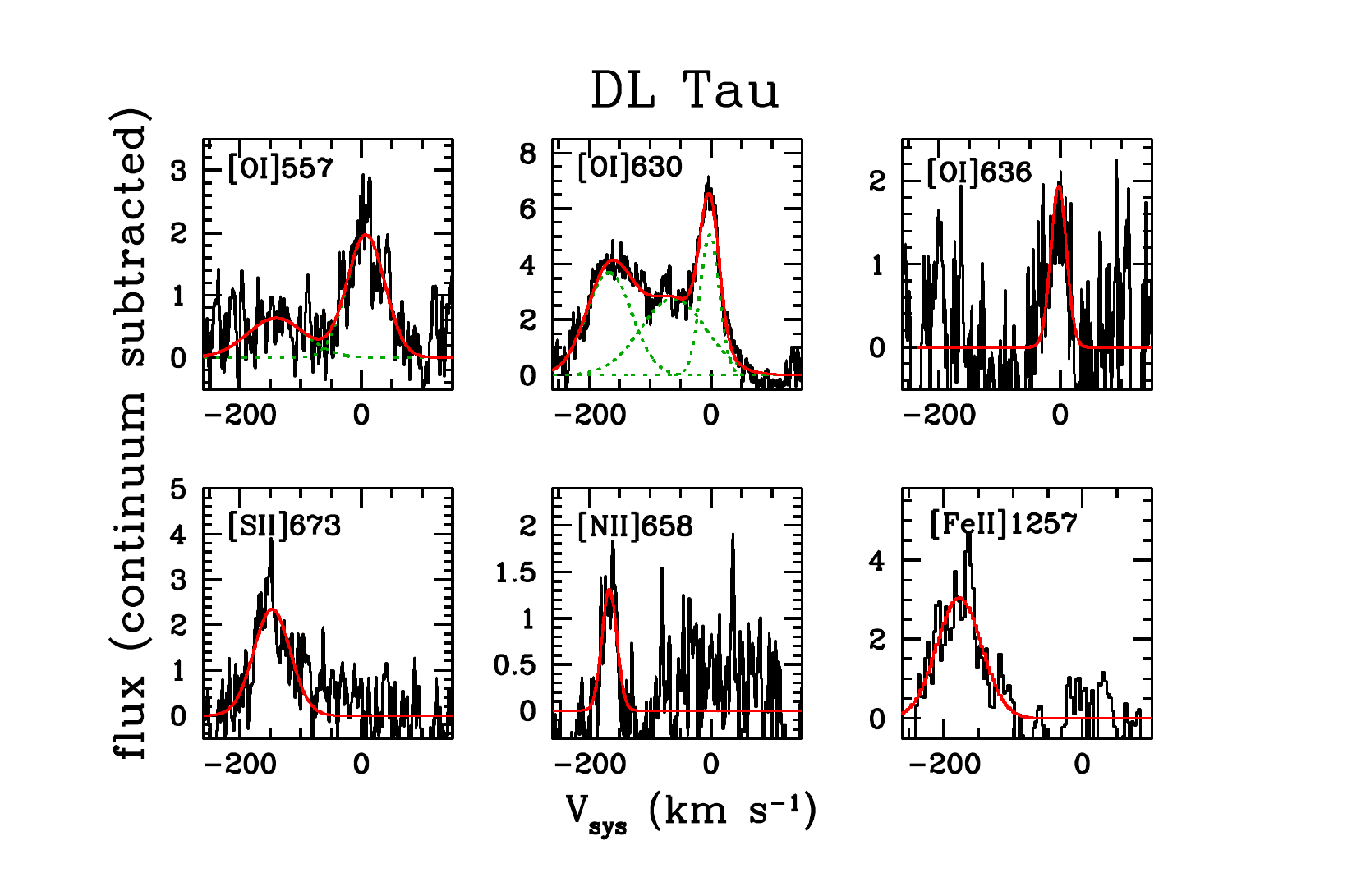}
  \caption{ \label{fig:fig3} As in Fig.\,\ref{fig:fig1} but for DL Tau.}
 \end{figure*}

\begin{figure*}
   \centering
   \includegraphics[width=18cm]{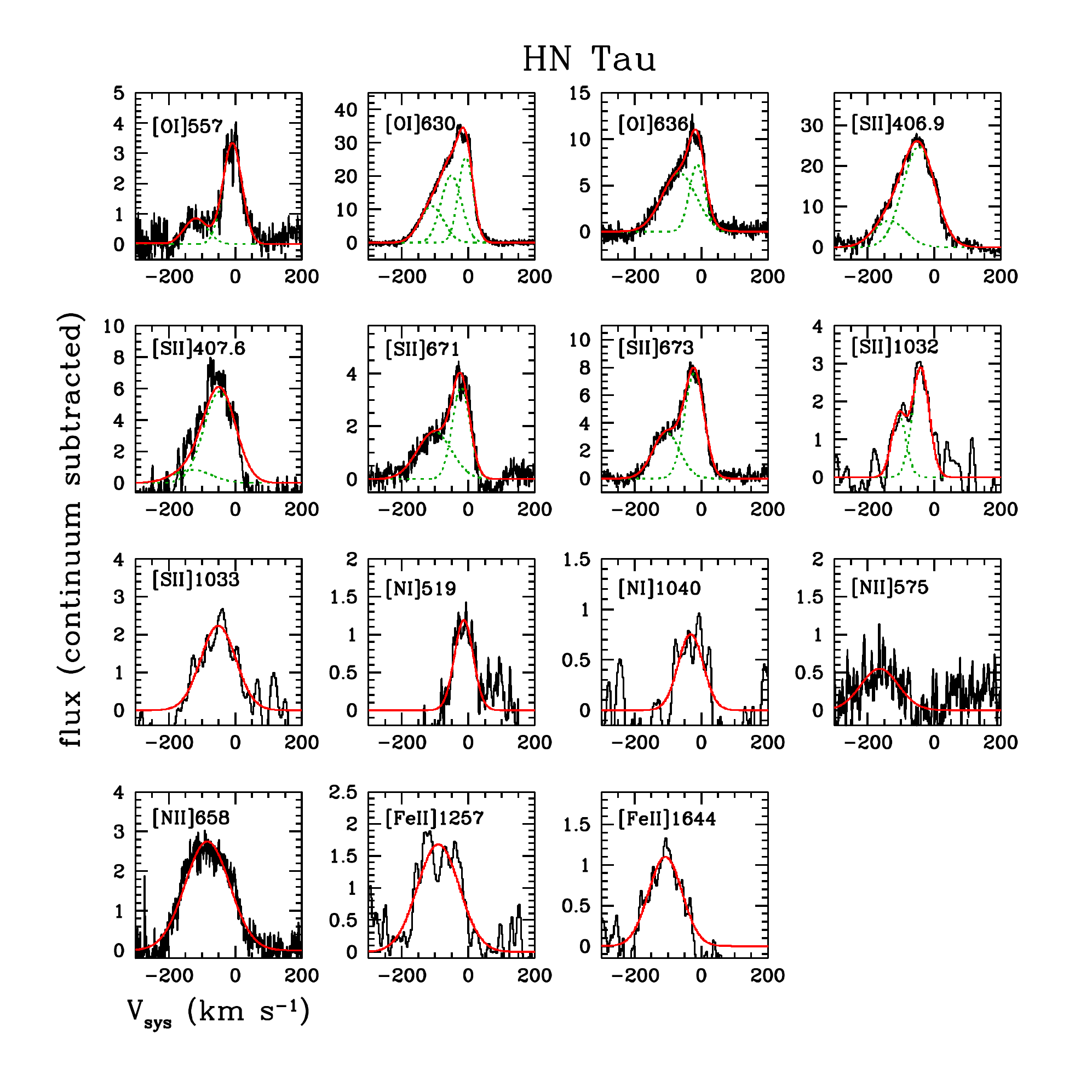}
   \caption{ \label{fig:fig4} As in Fig.\,\ref{fig:fig1} but for HN Tau.}
 \end{figure*}

\begin{figure*}
  \centering
   \includegraphics[width=18cm]{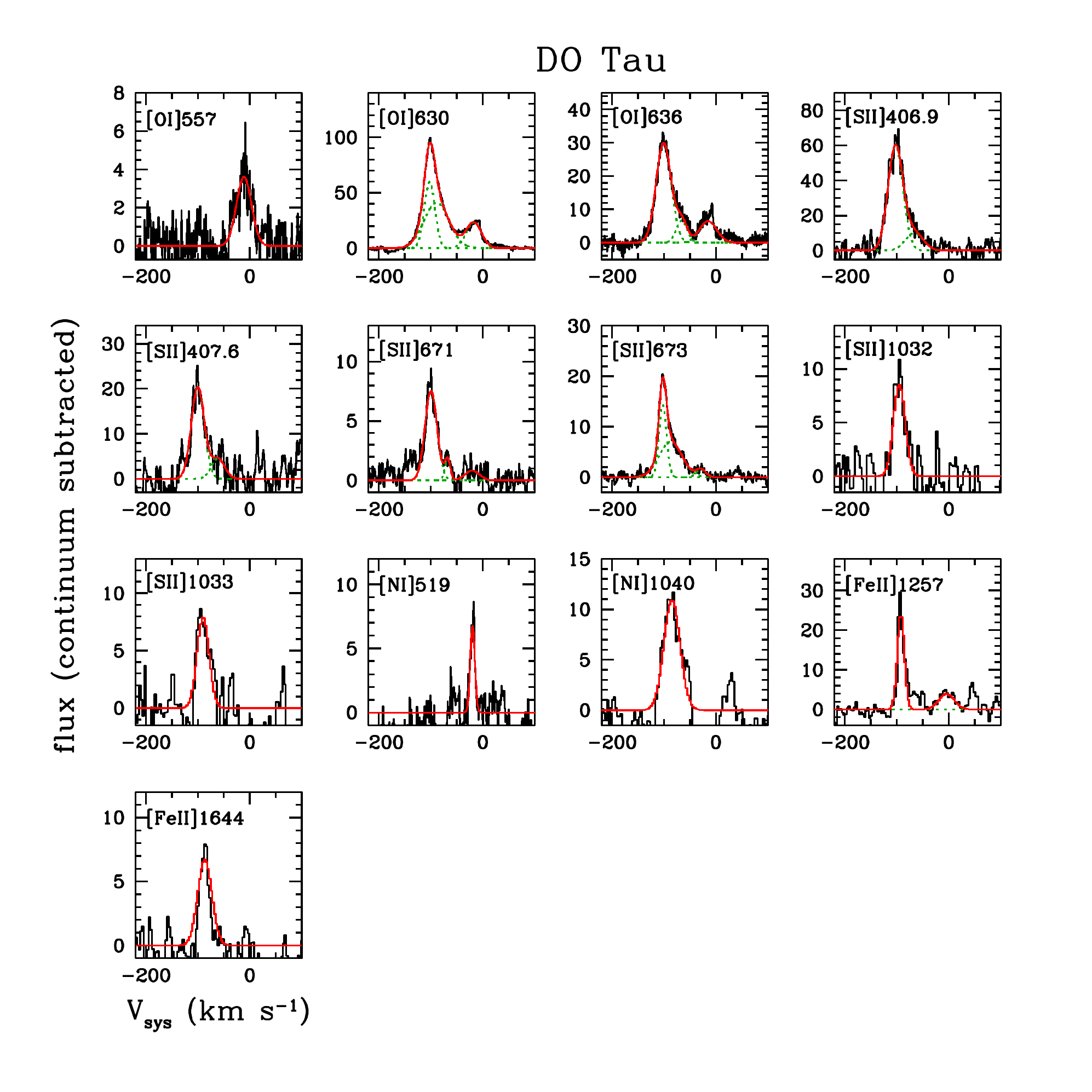}
   \caption{ \label{fig:fig5} As in Fig.\,\ref{fig:fig1} but for DO Tau.}
 \end{figure*}

\begin{figure*}
   \centering
   \includegraphics[width=18cm]{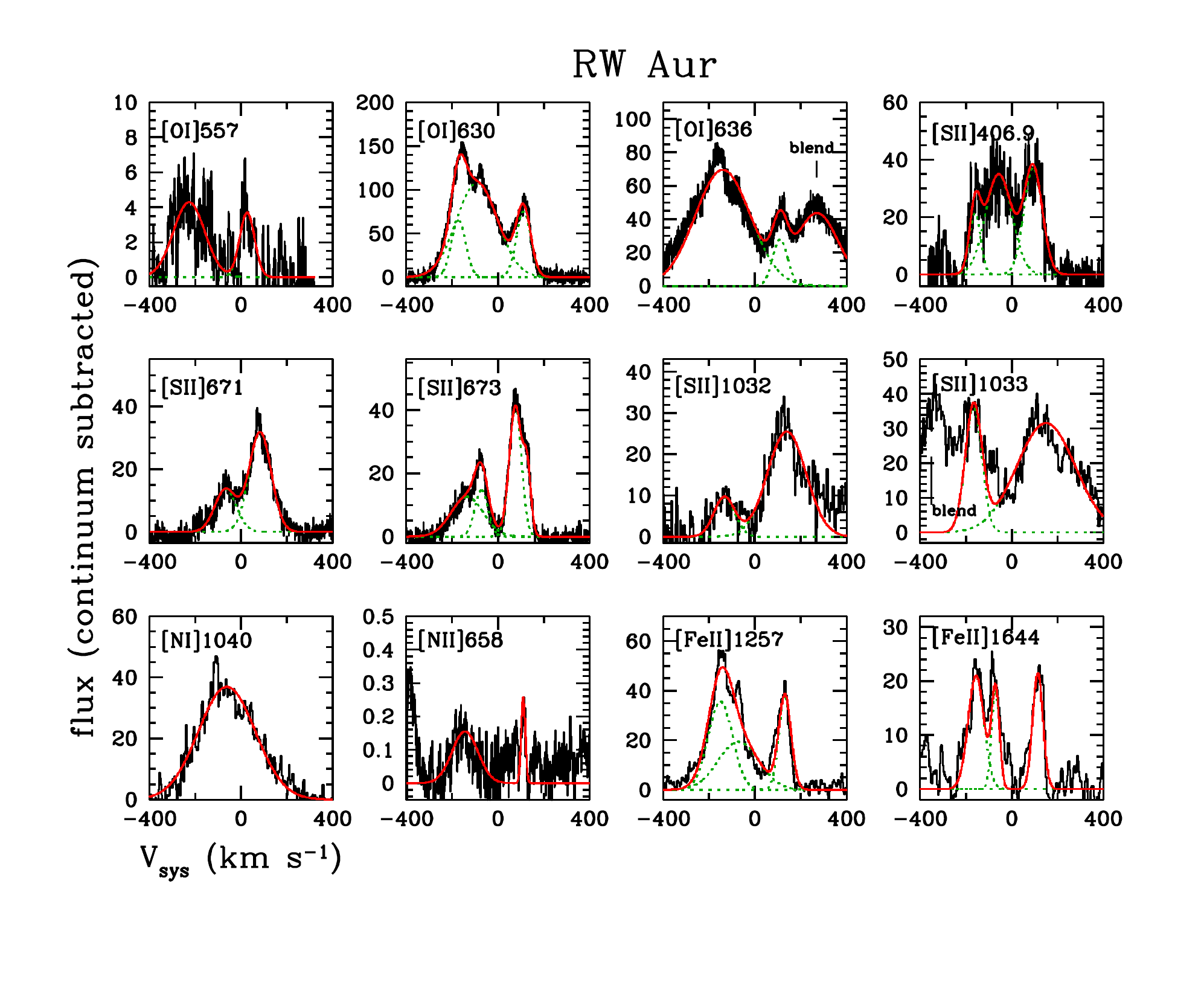}
   \caption{ \label{fig:fig6} As in Fig.\,\ref{fig:fig1} but for RW Aur.}
 \end{figure*}

The wide wavelength range covered by GIARPS allowed us to observe several forbidden lines spanning a large range of excitation and ionization conditions. In particular, we observed, in four out of the six sources, several [\oi], [\sii], [\n], [\nii], and [\feii] lines. A summary of the relevant atomic parameters of these lines is given in Table\,\ref{tab:tab3} while the continuum-subtracted spectral profiles are shown in 
Figs.\,\ref{fig:fig1}-\ref{fig:fig6}. 
Typically, the lines are blueshifted, although in several cases a red wing, extending up to a few tens of kilometres per second is also detected. An exception is RW Aur A (hereafter RW Aur), in which the redshifted component extends up to hundreds of kilometres per second.  
Here we provide a qualitative description of the line profiles, and a comparison with the literature observations obtained during the last decade especially for the [\oi]630 line.
\begin{itemize}
\item[-]{\bf RY Tau.} Only two lines ([\oi]630 and [\sii]673) are clearly detected in our spectrum (Fig.\,\ref{fig:fig1}). The [\oi]630 profile presents two blended components, a stronger LVC and a weaker blueshifted component at around\, $-$70 \kms\,, while [\sii]673 is detected only in the LVC. Both lines have been observed in 2010 by Chou et al. (2013). No significant variations can be recognized with respect to our spectra,  but the [\sii]673 line is barely resolved by Chou et al. and the two [\oi] components are more separated. No signature of the [\feii]1644 line, reported by Coffey et al. (2015), is present in our spectrum. Since the data of Coffey et al. are not flux-calibrated, we can not evaluate whether the [\feii] emission has dropped during recent years. Our 3$-\sigma$ upper limit of 5 10$^{-13}$ erg s$^{-1}$ cm$^{-2}$, derived from the rms around the line, is similar to the flux measured in the other jets of our sample (see Table\,\ref{tab:tab3}). This might indicate that the lack of detection is due to the poor line-to-continuum ratio around 1.6 $\mu$m, where the source is remarkably bright (see Table\,\ref{tab:tab2}), rather than to a real variation of the line.
\item[-]{\bf DG Tau}. This source has been extensively observed in recent decades. In 1998/1999, the blueshifted  [\oi]630 peaked at zero velocity with a wing extending up to  $-$450 km\,s$^{-1}$ (Lavalley-Fouquet et al. 2000, Bacciotti et al. 2000, Maurri et al. 2014). A secondary, less pronounced peak at a velocity of $v_{\rm p}\approx-$150 \kms\, was visible in the  spectrum of 2010 by Chou et al. (2013), and became more intense in 2012 (Iguchi and Itoh 2016). At that time the blue component was very bright also in the [\sii]406.9 profile. Another spectrum has been published by  Simon et al. (2016). The observation date is not specified, but the [\oi]630 profile shape closely resembles that of  Chou et al.  in 2010.
A progressive increase of the high-velocity blueshifted peak is confirmed by our observations (Fig.\,\ref{fig:fig2}). The  HVC  of [\oi]630  now shows a comparable intensity as the LVC  and extends up to $-$200 km\,s$^{-1}$. The HVC is also the main component of other [\feii], [\sii], and [\nii] lines. Unlike in 2012 (Iguchi and Itoh 2016), the HVC is now clearly detected in the [\oi]557 profile.
\item[-]{\bf DL Tau}. Six lines are detected in this spectrum (Fig.\,\ref{fig:fig3}). The brightest is [\oi]630, which is composed of an LVC and 
a weak HVC. [\oi]557  displays a similar profile, while another three lines, namely [\sii]673, [\nii]658, 
and [\feii]1257, show only the HVC
($v_{\rm p}$ $\approx$ $-$160 \kms\,). Simon et al. (2016) observed the [\oi]630 and [\sii]673
 lines. Their profiles are similar to the present ones, but the HVC of [\oi]630 is at the same level of intensity as the LVC. The [\oi]557 remained undetected in the Simon et al. spectrum.
\item[-]{\bf HN Tau}. In this spectrum we detect the [\oi], [\sii], [\n], [\nii], and [\feii] lines (Fig.\,\ref{fig:fig4}). All of them are blueshifted, with a wing extending up to $\approx$ $-$200 \kms\,. The lines reported in the recent literature are [\oi]630 and [\sii]673 by Simon et al. (2016), and 
[\oi]557,630 and [\sii]406.9 by Iguchi and Itoh (2016). The spectral profiles of these latter authors do not show any significant difference compared with the GIARPS data, likely indicating that no important physical or kinematic changes have occurred in the gas during the last decade.
\item[-]{\bf DO Tau}. Most of the lines detected with GIARPS show a blueshifted LVC and a HVC, with the latter ($v_{\rm p} \approx-$110 \kms\,) being the most prominent (Fig.\,\ref{fig:fig5}). Exceptions are [\oi]557 and [\n]519.8, detected in the LVC only. A comparison with literature data, namely the  [\oi]557/630, and [\sii]673 profiles of Simon et al. (2016), shows no significant variability compared with the GIARPS spectra.
\item[-]{\bf RW Aur}.  At variance with all the other sources, no LVC peak is detected. This is the only source for which we see a redshifted component (Fig,\,\ref{fig:fig6}). The latter is the most prominent feature in [\sii] lines, while [\oi] and [\feii] lines are brighter in the blue component. [\n]1040  appears as a single, broad (FWHM$\simeq$~300 km\,s$^{-1}$) and symmetric feature, which is blueshifted by a few tens of kilometres per second. In the red lobe, lines peak typically around $+$130 km s$^{-1}$, while in the blue lobe they peak at $\sim$ $-$150 km s$^{-1}$. 
In 2000, lines of the blue lobe peaked at  $\sim$ $-$200 km s$^{-1}$ (Woitas et al. 2005, Melnikov et al. 2009). In 2010 (Chou et al. 2013), the [\oi]630 line of RW Aur presented a flat shape, with a blue and red component of similar intensity. The blue component was just barely recognizable in the [\sii]673 line.
\end{itemize}

\begin{table*}
\caption{\label{tab:tab3} Relevant atomic parameters of the observed lines.}
\centering
\begin{tabular}{cccccccccc}
\hline\hline
Ion    & IP$_{\textrm{i-1}}-$IP$_\textrm{i}$&  Upper   &  Lower              &  $\lambda$  &  $T_\textrm{u}$  &  $T_\textrm{l}$   &   g$_\textrm{u}$A$_\textrm{ul}$    & $n_\textrm{cr}$  & Instr.    \\
%\hline
           &  (eV)              &  Level   &  Level              &  (nm)       &  (K)       &  (K)      &  (s$^{-1}$)   & (cm$^{-3}$)    &           \\
\hline\hline
\oi        & 0$-$13.62          &  $^1$S$_0$     & $^1$D$_2$     &  557.7     &  48619.93  &  22830.29  & 1.26e+0      &   9.2e+7     &  H    \\
\oi        &                    &  $^1$D$_2$     & $^3$P$_2$     &  630.0     &  22830.29  &  0.00      & 3.25e-2      &   1.6e+6     &  H    \\
\oi        &                    &  $^1$D$_2$     & $^3$P$_1$     &  636.4     &  22830.29  &  227.71    & 2.10e-3      &   1.6e+6     &  H    \\
\sii       & 10.36$-$23.34      &  $^2$P$_{3/2}$ & $^4$S$_{3/2}$ &  406.9     &  35352.92  &  0.00      & 7.70e-1      &   1.9e+6     &  H    \\
\sii       &                    &  $^2$P$_{1/2}$ & $^4$S$_{3/2}$ &  407.6     &  35285.72  &  0.00      & 1.55e-1      &   2.6e+6     &  H    \\
\sii       &                    &  $^2$D$_{5/2}$ & $^4$S$_{3/2}$ &  671.6     &  21415.78  &  0.00      & 1.21e-3      &   1.7e+3     &  H    \\
\sii       &                    &  $^2$D$_{3/2}$ & $^4$S$_{3/2}$ &  673.1     &  21370.04  &  0.00      & 2.27e-3      &   1.6e+4     &  H    \\
\sii       &                    &  $^2$P$_{3/2}$ & $^2$D$_{5/2}$ &  1032.3    &  35352.92  & 21415.78   & 6.25e-1      &   1.9e+6     &  G    \\
\sii       &                    &  $^2$P$_{1/2}$ & $^2$D$_{3/2}$ &  1033.9    &  35285.72  & 21370.04   & 2.86e-1      &   2.6e+6     &  G    \\
\n         & 0$-$14.53          &  $^2$D$_{3/2}$ & $^4$S$_{3/2}$ &  519.8     &  27672.22  &  0.00      & 8.12e-5      &   9.5e+3     &  H    \\
\n$^a$     &                    &  $^2$P$_{3/2}$ & $^2$D$_{5/2}$ &  1040.06   &  41493.28  & 27659.68   & 2.45e-1      &   3.0e+6     &  G    \\
\n$^a$     &                    &  $^2$P$_{1/2}$ & $^2$D$_{5/2}$ &  1040.10   &  41492.72  & 27659.68   & 6.89e-2      &   3.1e+6     &  G    \\
\nii       & 14.53$-$29.60      &  $^1$S$_0$     & $^1$D$_2$     &  575.5     &  47031.60  & 22036.55   & 1.14e+0      &   1.2e+7     &  H    \\          
\nii       &                    &  $^1$D$_2$     & $^3$P$_2$     &  658.3     &  22036.55  & 188.19     & 1.45e-2      &   8.0e+4     &  H    \\
\feii      & 7.90$-$ 16.19      & a$^4$D$_{7/2}$ &a$^6$D$_{9/2}$ &  1257.0    &  11445.92  &  0.00      & 4.48e-2      &   6.8e+4     &  G    \\
\feii      &                    & a$^4$D$_{7/2}$ &a$^4$F$_{9/2}$ &  1644.0    &  11445.92  & 2694.25    & 1.52e-2      &   6.8e+4     &  G    \\
\hline
\end{tabular}
\tablefoot{We give $\lambda_{\rm air}$ and $\lambda_{\rm vac}$ for lines in the HARPS-N (H) and GIANO-B (G) range, respectively. The critical density is computed at \Te\,= 10\,000 K.\\
$^a$Blended lines.}
\end{table*}

\subsection{Kinematic parameters}\label{sec:sec3.2}

%-------------------------------------- Figura 1
%-------------------------------------- Figura 2
%
%-------------------------------------- Figura 3
%
%-------------------------------------- Figura 4
%
%-------------------------------------- Figura 5
%
%-------------------------------------- Figura 6

To get more quantitative information on the kinematic properties of the detected lines, we used the 
tool {\sc splot} within the IRAF package to deconvolve each profile in multiple Gaussians,  deriving  the peak intensity and velocity ($I_{\rm p}$ and $v_{\rm p}$)  and the FWHM of all components. In all objects the brightest line is the [\oi]630. We use the peak of this line to empirically separate the velocity components, namely the low-velocity component  (LVC\,: $-$40 \lapprox\, $v_{[\oi]630}$\,\lapprox\, 40 km\,s$^{-1}$), the  medium-velocity component (MVC\,: $-$100 \lapprox\, $v_{[\oi]630}$\, \lapprox\,$-$40 km\,s$^{-1}$), 
and the high-velocity component (HVC\,: $v_{[\oi]630}$\, \lapprox\,$-$100 km\,s$^{-1}$). The same components are attributed to the other lines from the visual comparison of their profile with that of the [\oi]630 line. 
 
\begin{figure*}
   \centering
   \includegraphics[width=14cm]{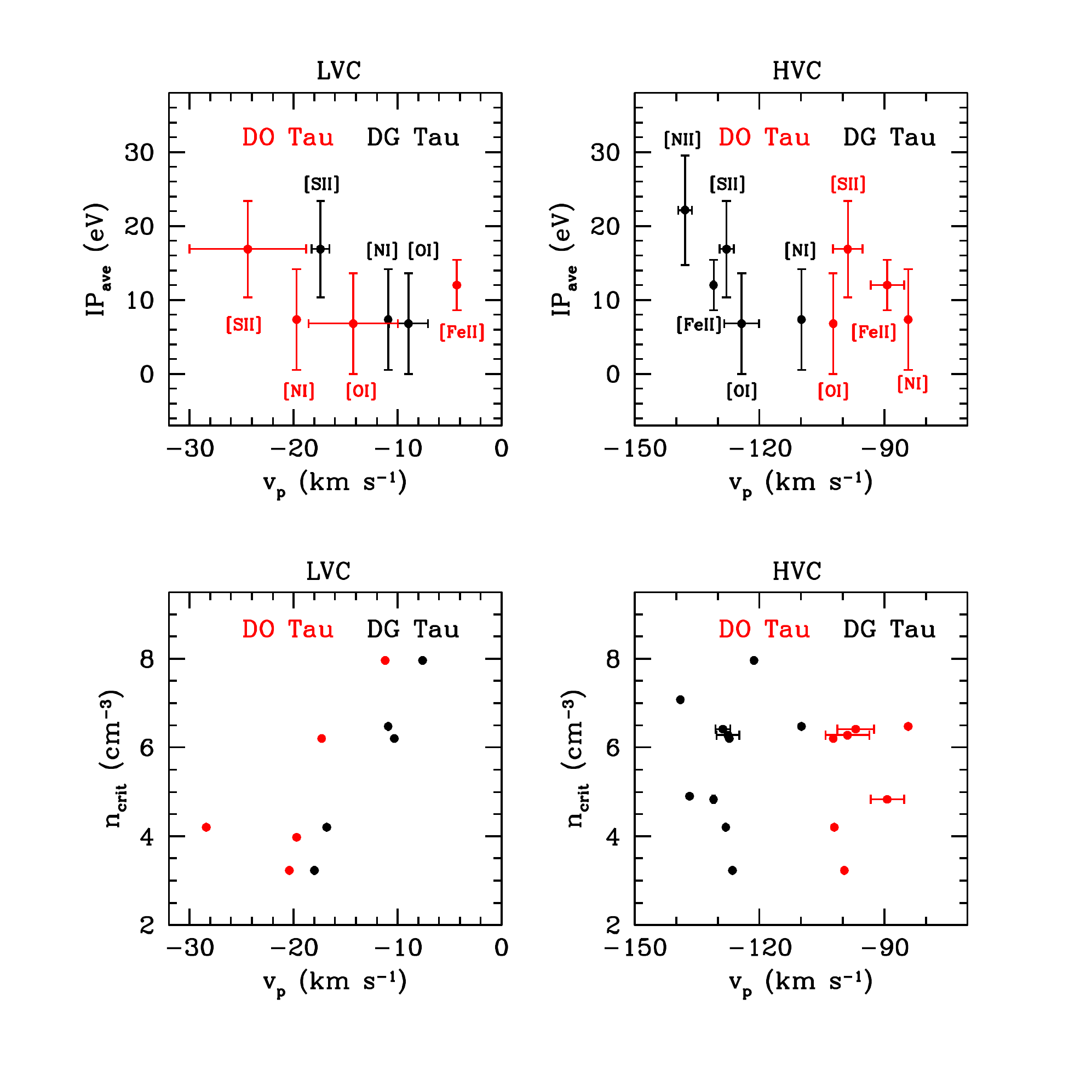}
   \caption{ \label{fig:fig7} {\it Upper panels}: IP$_\textrm{ave}$ vs. $v_\textrm{p}$, where IP$_\textrm{ave}$=(IP$_\textrm{i-1}$ +IP$_\textrm{i}$)/2 and IP$_\textrm{i}$ is the ionization potential of the ionic stage $i$. The left and right panels
   refer to the LVC and the HVC, respectively. Data points of DO Tau and DG Tau are plotted with red and black dots, respectively. The error bars  in the x-axis represent the dispersion between the $v_\textrm{p}$
   values of lines of the same ionic species, while those in the y-axis are (IP$_\textrm{i}$-IP$_\textrm{i-1}$)/2.  {\it Lower panels}:  Critical density ($n_\textrm{crit}$)
   of the upper level of the transition vs. $v_\textrm{p}$. Symbols have the same meaning as in the upper panels. }
 \end{figure*}

The results are given in Table\,\ref{tab:tab4}. At a first glimpse, we note a relation between the presence of an LVC or a HVC and the ionization or excitation degree of the emission lines.  Indeed, an LVC is detected primarily in low-excitation lines of neutral species ([\oi]\,, 
[\n]\,), and in [\sii]671/673.  The HVC is instead detected (or much stronger than the LVC) in [\nii]\,, [\feii]\,, and  [\sii]\, lines of higher excitation energy (e.g., [\sii]406.9). Besides, both $v_{\rm p}$ and FWHM vary significantly among the lines of the same species. These results indicate that a strong dependence exists between kinematic and excitation properties of the gas in the flow. Here we qualitatively investigate such a relation relying on observational data only, while in the following (Sect.\ref{sec:sec4}) a detailed diagnostic analysis is carried out based on our excitation and ionization model. 
We consider as examples the DG Tau and DO Tau jets, which display lines with well-separated LVC and HVC.
In Fig.\,\ref{fig:fig7} (upper panels), IP$_\textrm{ave}$\,=\, (IP$_\textrm{i-1}$ +IP$_\textrm{i}$)/2, namely 
the average between the ionization potentials of the ionic stages $i$-1 and $i$, is plotted against the peak velocity of the LVC and the HVC. IP$_{\rm ave}$ gives an idea of the energy range in which a given element is mostly found in the ionic stage $i$.
  The error bars in the x-axis represent the dispersion between the $v_\textrm{p}$ values of lines of the same ionic species, while those in the y-axis correspond to (IP$_\textrm{i}$ -IP$_\textrm{i-1}$)/2.  No correlation between IP$_\textrm{ave}$  and $v_\textrm{p}$ is found for 
 DO Tau, while for DG Tau,  IP$_\textrm{ave}$  roughly increases with $v_\textrm{p}$ of both components, especially for the HVC. This result reveals in advance the more quantitative conclusion of Sect.\,\ref{sec:sec4}, namely that the ionization degree in DG Tau has a clear dependence on velocity, while it remains almost constant in all the velocity components of DO Tau.

The relation between gas density and velocity is displayed in the lower panels, where we plot the critical density ($n_\textrm{crit}$) of the level from which the line originates against $v_\textrm{p}$. The critical density is calculated within our nonlocal thermal equilibrium (NLTE) code (see Appendix\,\ref{sec:sec4.1.1}), assuming an electron temperature of 10\,000 K. Interestingly, we note a decrease of $n_\textrm{crit}$ with  $v_\textrm{p}$ for the LVC of both sources, while no correlation exists for the HVC. This effect, firstly noted by Hartigan et al. (1995) for the [\oi]630 and [\sii]673 lines, is consistent with the picture in which lines with higher critical densities originate closer to the disk surface, where the wind traced by the LVC accelerates away from the disk. 

%
%-------------------------------------- Figura 7

\section{Analysis}\label{sec:sec4}
\subsection{Diagnostic analysis}\label{sec:sec4.1}

So far, several studies have addressed the physical properties of the gas in jets to constrain the excitation and ionization mechanisms (see e.g., Alexander et al. 2014 and references therein). In most of those works, the plasma physical conditions have been inferred mainly using a diagnostic analysis based on the ratios of the [\oi]630, [\sii]671,673, and [\nii]658 lines, as originally proposed by  Bacciotti and Eisl\"offel (1999). This technique has the advantage of using bright lines sensitive to variations of electron density and ionization fraction and close in wavelength, thus minimizing extinction effects. However, all the considered lines have similar excitation energy, and therefore they are not very sensitive to the temperature. In addition, the derivation of the electron density relies on the [\sii]671/[\sii]673 ratio, which is sensitive only to densities less than $\sim$ 10$^4$  cm$^{-3}$.

A multi-line approach, like the one we adopt here and originally proposed by Hartigan and Morse (2007), provides a better constraint on the beam-averaged physical parameters, minimizing the bias due to the selection of specific lines. 
In particular, in the present work we determine the physical conditions of the gas by comparing the observed flux ratios with the predictions of our excitation and ionization model (Giannini et al. 2013, 2015), the main features of which are briefly described in Appendix\,\ref{sec:sec4.1.1}.

\subsubsection{Physical parameters versus velocity channels} \label{sec:sec4.1.2}

\begin{table*}
\footnotesize
\caption{\label{tab:tab4} Line kinematic parameters. Peak velocity ($v_\textrm{p}$) and FWHM are in \kms, while peak intensity ($I_\textrm{p}$) is in erg s$^{-1}$ cm$^{-2}$ \AA$^{-1}$.}
\centering
\begin{tabular}{cccc|ccc|ccc}
\hline\hline
%\multicolumn{5}{c}{RY Tau}   \\ 
                    & \multicolumn{3}{c|}{HVC}                       &\multicolumn{3}{c|}{MVC}                    & \multicolumn{3}{c}{LVC}                    \\
%\hline                                                                                                                                                      
Source/Line         &  $v_\textrm{p}$      &    FWHM      &    $I_\textrm{p}$          & $v_\textrm{p}$      & FWHM           &   $I_\textrm{p}$      & $v_\textrm{p}$         & FWHM       &   $I_\textrm{p}$        \\
\hline                                                                                                                                                       
\bf{RY Tau}         &             &              &                   &            &                &              &               &            &               \\    
%\hline                                                                                                                                                      
$[\oi]$\,630.0      &    -        &    -         &      -            & -66.9      & 57.0           &  3.12e-14    & -20.4         & 39.0       &   4.24e-14    \\
$[\sii]$\,673.1     &    -        &    -         &      -            &   -        &  -             &   -          & -27.0         & 49.0       &   7.20e-15    \\
\hline                                                                                                                                                       
\bf{DG Tau}         &             &              &                   &            &                &              &              &             &               \\  
$[\oi]$\,557.7      & -121.3      &    91.6      &   7.24e-15        &   -        &   -            &   -          &  -7.6        & 23.7        &  2.88e-14     \\
$[\oi]$\,630.0      & -127.3      &    54.0      &   2.53e-13        &   -53.7    & 74.3           &  1.39e-13    &  -10.3       & 22.1        &  1.99e-13     \\
$[\oi]$\,636.4      & -127.6      &    56.5      &   8.22e-14        &   -51.5    & 77.7           &  4.47e-14    &  -9.7        & 21.3        &  6.44e-14     \\
$[\sii]$\,406.9     & -129.5      &    50.2      &   9.55e-14        &   -57.6    & 97.2           &  6.52e-14    &  -           &  -          &     -         \\
$[\sii]$\,407.6     & -130.1      &    46.6      &   1.90e-14        &   -63.1    & 164.4          &  1.96e-14    &  -           &  -          &     -         \\
$[\sii]$\,671.6     & -126.5      &    52.1      &   1.69e-14        &   -48.5    & 70.9           &  1.50e-14    &  -18.0       & 24.8        &  1.61e-14     \\
$[\sii]$\,673.1     & -128.1      &    47.2      &   3.98e-14        &   -52.3    & 70.0           &  3.29e-14    &  -16.8       & 28.5        &  3.98e-14     \\
$[\sii]$\,1032.3    & -125.7      &    53.6      &   4.34e-14        &   -49.4    & 79.8           &  2.03e-14    &  -           &  -          &      -        \\
$[\sii]$\,1033.9    & -127.5      &    68.8      &   3.03e-14        &   -47.8    & 60.2           &  1.25e-14    &  -           &  -          &      -        \\
$[\n]$\,1040.06/1040.10& -109.9   &    58.0      &   2.79e-14        &   -59.5    & 37.5           &  1.03e-14    &  -10.9       & 62.0        &  2.00e-14     \\
$[\nii]$\,575.5     & -139.1      &    65.6      &   3.75e-15        &    -       &  -             &      -       &  -           &  -          &      -        \\
$[\nii]$\,658.3     & -136.8      &    43.8      &   3.73e-14        &   -100.0   & 133.6          &  1.15e-14    &  -           &  -          &      -        \\
$[\feii]$\,1257.0   & -130.9      &    50.7      &   2.45e-14        &    -71.8   & 98.9           &  2.81e-15    &  -           &  -          &      -        \\
$[\feii]$\,1644.0   & -131.2      &    47.7      &   1.51e-14        &    -65.8   & 39.5           &  3.16e-15    &  -           &  -          &      -        \\
\hline                                                                                                                                                       
\bf{DL Tau}         &             &              &                   &            &                &              &              &             &               \\
$[\oi]$\,557.7      & -139.9      &   111.4      &   6.32e-16        &    -       &  -             &   -          &  7.5         & 73.0        &  1.97e-15     \\
$[\oi]$\,630.0      & -166.89     &   87.6       &   3.69e-15        &  -61.9     & 126.3          &  2.76e-15    & -1.9         & 37.0        &  4.79e-15     \\
$[\oi]$\,636.4      &   -         &    -         &      -            &   -        &  -             &     -        & -2.6         & 30.5        &  1.93e-15     \\
$[\sii]$\,673.1     & -146.7      &    71.0      &   2.34e-15        &    -       &  -             &     -        &  -           &  -          &      -        \\
$[\nii]$\,658.3     & -166.6      &    28.6      &   1.30e-15        &    -       &  -             &     -        &  -           &  -          &      -        \\
$[\feii]$\,1257.0   & -177.4      &    76.3      &   3.04e-15        &    -       &  -             &     -        &  -           &  -          &     -         \\
\hline                                                                                                                                                       
\bf{HN Tau}         &             &              &                   &            &                &              &              &             &               \\ 
$[\oi]$\,557.7      & -118.5      &    81.0      &   8.52e-16        &    -       &  -             &      -       &   -8.6       & 67.2        &   3.34e-15    \\
$[\oi]$\,630.0      & -109.9      &    84.3      &   1.10e-14        &   -51.1    & 71.8           &  2.02e-14    &   -7.4       & 52.6        &   2.56e-14    \\
$[\oi]$\,636.4      &  -          &     -        &    -              &   -66.8    & 120.6          &  6.34e-15    &   -12.4      & 55.0        &   7.19e-15    \\
$[\sii]$\,406.9     & -134.5      &   113.3      &   4.52e-15        &   -46.3    & 114.3          &  2.48e-14    &    -         &  -          &     -         \\
$[\sii]$\,407.6     & -129.8      &   133.7      &   8.43e-16        &   -46.2    & 114.0          &  5.81e-15    &    -         &  -          &      -        \\
$[\sii]$\,671.6     & -101.8      &   121.4      &   1.79e-15        &     -      &  -             &     -        &  -21.7       & 60.1        &   3.47e-15    \\
$[\sii]$\,673.1     & -105.2      &    95.0      &   3.35e-15        &     -      &  -             &    -         &  -20.8       & 29.4        &   7.60e-15    \\
$[\sii]$\,1032.3    & -126.6      &    49.6      &   1.63e-15        &   -60.7    & 59.4           &  2.89e-15    &    -         &   -         &      -        \\
$[\sii]$\,1033.9    &   -         &     -        &     -             &   -71.0    & 124.3          &  2.83e-15    &  -           &  -          &     -         \\
$[\n]$\,519.8       &   -         &     -        &    -              &    -       &  -             &    -         &  -13.1       & 66.6        &    1.19e-15   \\
$[\n]$\,1040.06/1040.10&   -      &     -        &     -             &   -51.8    &  81.7          &  7.53e-16    &    -         &  -          &     -         \\
$[\nii]$\,575.5     & -139.1      &    65.6      &    5.45e-16       &    .       &  -             &      -       &  -           &  -          &        -      \\
$[\nii]$\,658.3     &   -         &     -        &      -            &   -83.9    & 157.7          &  2.75e-15    &    -         &  -          &     -         \\
$[\feii]$\,1257.0   & -109.7      &    148.1     &    1.68e-15       &    -       &   -            &     -        &    -         &  -          &     -         \\
$[\feii]$\,1644.0   & -107.9      &    118.0     &    1.10e-15       &    -       &   -            &     -        &    -         &  -          &     -         \\
\hline                                                                                                                                                       
\bf{DO Tau}         &             &              &                   &            &                &              &              &             &               \\ 
$[\oi]$\,557.7      &  -          &      -       &     -             &    -       &   -            &     -        &  -11.2       & 33.9        &    3.61e-15   \\
$[\oi]$\,630.0      & -102.3      &     24.1     &   5.93e-14        &  -89.0     &   56.5         &  4.10e-14    &  -17.3       & 34.6        &    2.31e-14   \\
$[\oi]$\,636.4      & -100.8      &     32.6     &   2.98e-14        &  -65.6     &   29.7         &  6.58e-15    &  -14.4       & 33.4        &    6.44e-15   \\
$[\sii]$\,406.9     & -102.6      &     34.0     &   3.94e-14        &  -66.1     &   40.3         &  1.04e-14    &   -          &  -          &      -        \\
$[\sii]$\,407.6     & -100.0      &     29.0     &   2.04e-14        &  -61.2     &   30.1         &  4.50e-15    &    -         &  -          &      -        \\
$[\sii]$\,671.6     & -99.6       &     27.3     &   8.73e-15        &  -67.7     &   12.3         &  1.78e-15    &  -20.4       & 32.7        &    8.28e-16   \\
$[\sii]$\,673.1     & -102.0      &     17.1     &   1.43e-14        &  -86.8     &   51.7         &  6.97e-15    &  -28.4       & 21.0        &    1.52e-15   \\
$[\sii]$\,1032.3    & -95.1       &     26.5     &   8.49e-15        &    -       &  -             &     -        &   -          &  -          &      -        \\
$[\sii]$\,1033.9    & -93.8       &     27.0     &   7.89e-15        &    -       &  -             &     -        &   -          &  -          &      -        \\
$[\n]$\,519.8       &  -          &     -        &     -             &    -       &   -            &     -        &  -19.7       & 10.4        &    6.73e-15   \\
$[\n]$\,1040.06/1040.10&  -84.3   &     36.6     &   1.09e-14        &    -       &   -            &     -        &   -          &  -          &       -       \\
$[\feii]$\,1257.0   &  -92.2      &     16.0     &   2.40e-14        &    -       &   -            &     -        &  -4.3        & 35.1        &    3.90e-15   \\
$[\feii]$\,1644.0   &  -86.5      &     32.1     &   6.69e-15        &    -       &   -            &     -        &   -          &  -          &     -         \\          
\hline                                                                                                                                                       
\bf{RW Aur}         &             &              &                   &            &                &              &              &             &                \\ 
$[\oi]$\,557.7      &  -226.5     &   162.1      &   4.29e-15        &     26.6   & 76.2           &   3.69e-15   &    -         &  -          &     -          \\
$[\oi]$\,630.0      &  -172.1     &    73.9      &   6.50e-14        &    -90.9   & 211.3          &   1.08e-13   &   111.6      & 74.0        &    7.55e-14    \\
$[\oi]$\,636.4      &  -161.3     &   294.8      &   6.96e-14        &     -      & -              &     -        &   109.8      & 71.8        &    2.78e-14    \\
$[\sii]$\,406.9     &  -159.6     &    55.5      &   2.14e-14        &    -57.0   & 134.9          &   3.50e-14   &    93.3      & 94.3        &    3.72e-14    \\
$[\sii]$\,671.6     &   -         &    -         &     -             &    -67.4   & 110.7          &   1.34e-14   &    83.2      & 113.0       &    3.19e-14    \\
$[\sii]$\,673.1     &  -134.0     &   164.5      &   1.26e-14        & -72.1/76.0 & 48.8/62.0      &1.46/4.11e-14 &   127.3      & 39.3        &    2.02e-14    \\
$[\sii]$\,1032.3    &  -132.9     &   107.5      &   9.56e-15        &      -     &  -             &     -        &   139.1      & 197.9       &    2.55e-14    \\
$[\sii]$\,1033.9    &  -166.0     &    85.8      &   3.59e-14        &      -     &  -             &     -        &   149.3      & 127.7       &    3.15e-14    \\
$[\n]$\,1040.06/1040.10&  -       &    -         &       -           &    -67.7   & 291.6          &    3.68e-14  &   -          &    -        &      -         \\
$[\nii]$\,658.3     & -152.3      &    61.3      &   1.54e-16        &     -      &  -             &     -        &   110.6      & 10.4        &    2.57e-16    \\
$[\feii]$\,1257.0  &  -150.4      &   125.4      &   3.57e-14        &    -72.3   & 207.7          &    1.94e-14  &   131.6      & 63.2        &    3.73e-14    \\
$[\feii]$\,1644.0  &  -155.3      &    81.6      &   2.09e-14        &    -70.7   & 25.5           &    1.84e-14  &   113.6      & 53.3        &    2.14e-14    \\                     
\hline\hline                                                                                                    
\end{tabular}
\end{table*}

The diagnostic analysis has been applied to DG Tau, HN Tau, DO Tau, and RW Aur, where a significant number of lines with a high S/N ratio in all the velocity components have been detected. In these sources, as shown in Table\,\ref{tab:tab4}, the kinematic parameters of lines within each velocity component may significantly differ from each other, with variations of more than 20 km\,s$^{-1}$ in $v_{\rm p}$ and larger than a factor of two in FWHM. To follow such variations, rather than integrating over the low-, medium-, and high-velocity components, we summed the line flux inside velocity channels.
The step in velocity has been chosen wide enough to maintain a high S/N ratio inside each channel, namely 20 km\,s$^{-1}$ in DG Tau and HN Tau, and 40/50 km\,s$^{-1}$ in DO Tau and RW Aur, respectively.

Flux ratios were then computed taking the [\oi]630 line as a reference, which is bright in all objects.  According to the line list of Table\,\ref{tab:tab4}, from nine to twelve flux ratios can be used in the four sources (flux ratios involving [\feii] lines are not considered at this step; see Sect.\,\ref{sec:sec4.2}). We accepted  flux ratios with S/N $\ge$ 5 as detections and assigned a 3-$\sigma$ upper limit to the others. 

The free parameters of our model are the electron temperature and density (\Te\,, \dens\,), the fractional ionization (\xe), and the visual extinction (\av\,), which is allowed to vary between 0 and 5 mag in steps of 0.5 mag. The best fit is then found recursively by applying a $\chi^2$ minimization to the differences between theoretical and de-reddened flux ratios. The upper limits are not considered in the fit but we verified  their
consistency {\it a posteriori}  with the best-fit model. 

\begin{figure*}[h!]
   \centering
   \includegraphics[width=18cm]{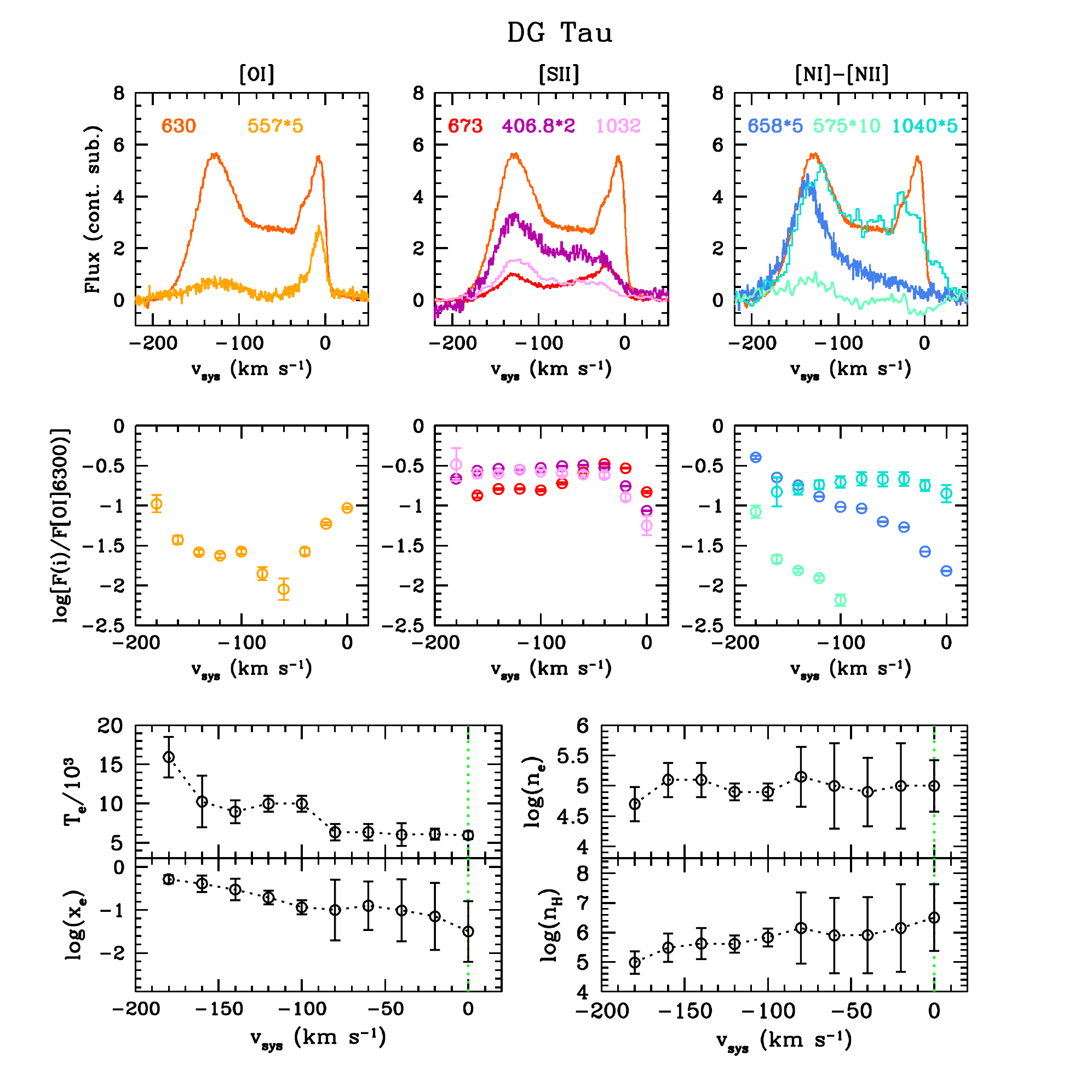}
   \caption{ \label{fig:fig10} {\it Upper panels}: Examples of the continuum-subtracted lines of DG Tau used in the fitting procedure. 
    Different lines are depicted with different colors according to the labels. For a better visualization, some fluxes have been multiplied by a constant as indicated. The [\oi]630 profile, which is taken as a reference in the fitting, is shown for comparison in each panel. {\it Middle panels}: De-reddened flux ratios with respect to [\oi]630 computed in bins of 20 km\,s$^{-1}$. Colors have the same meaning as in the top panels. {\it Bottom panels}: Gas parameters (\Te, \dens, \xe, \nH) plotted as a function of the bins of velocity. 
     The error bars correspond to models whose  $\chi^2$ is up to 30\% higher than that of the best fit. The green vertical dotted line marks the position of the rest velocity.}
 \end{figure*}

\begin{figure*}
   \centering
   \includegraphics[width=18cm]{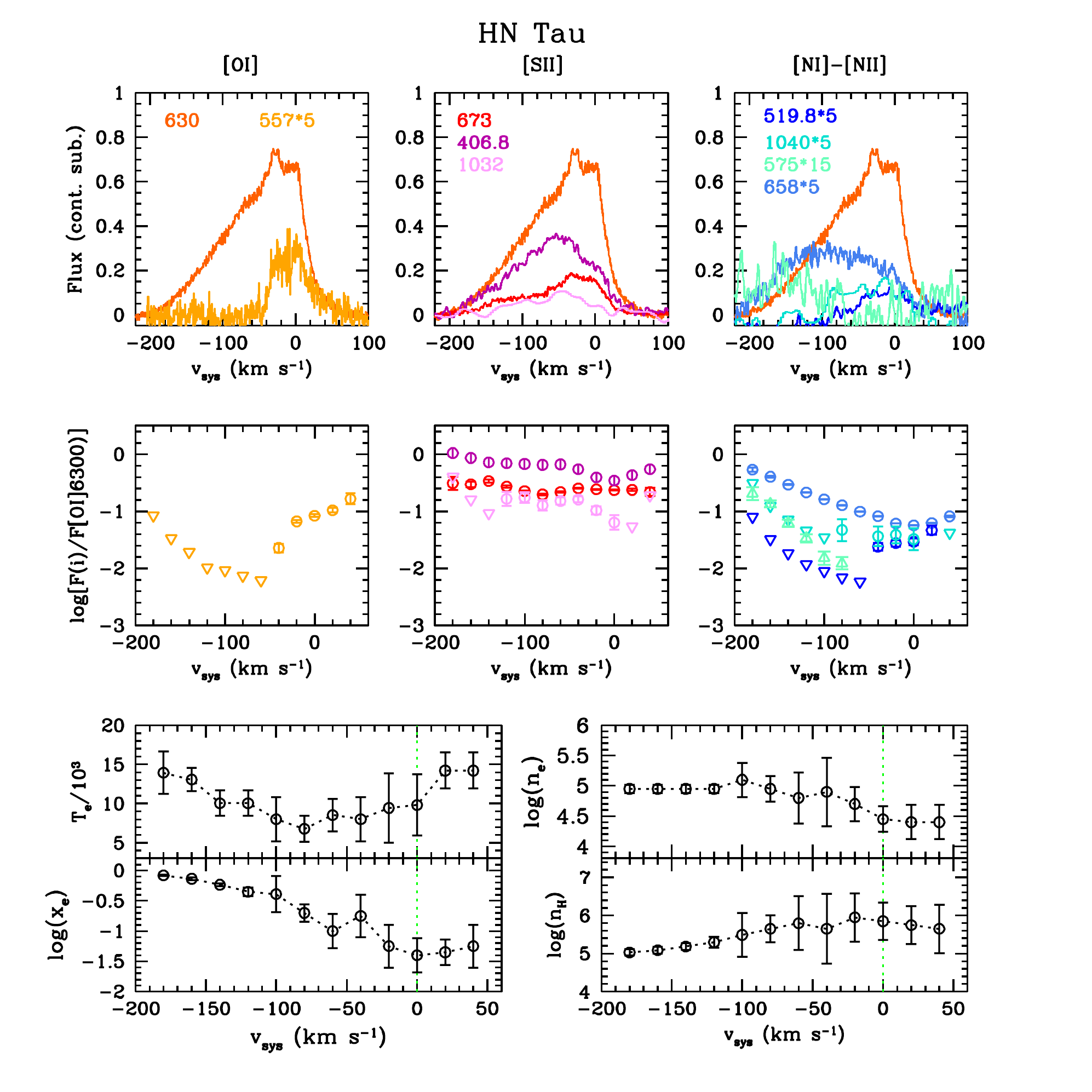}
   \caption{ \label{fig:fig11} As in Fig.\,\ref{fig:fig10} but for HN Tau. In the middle panels, reverse triangles are 3-$\sigma$ upper limits.}
 \end{figure*}

\begin{figure*}
   \centering
   \includegraphics[width=18cm]{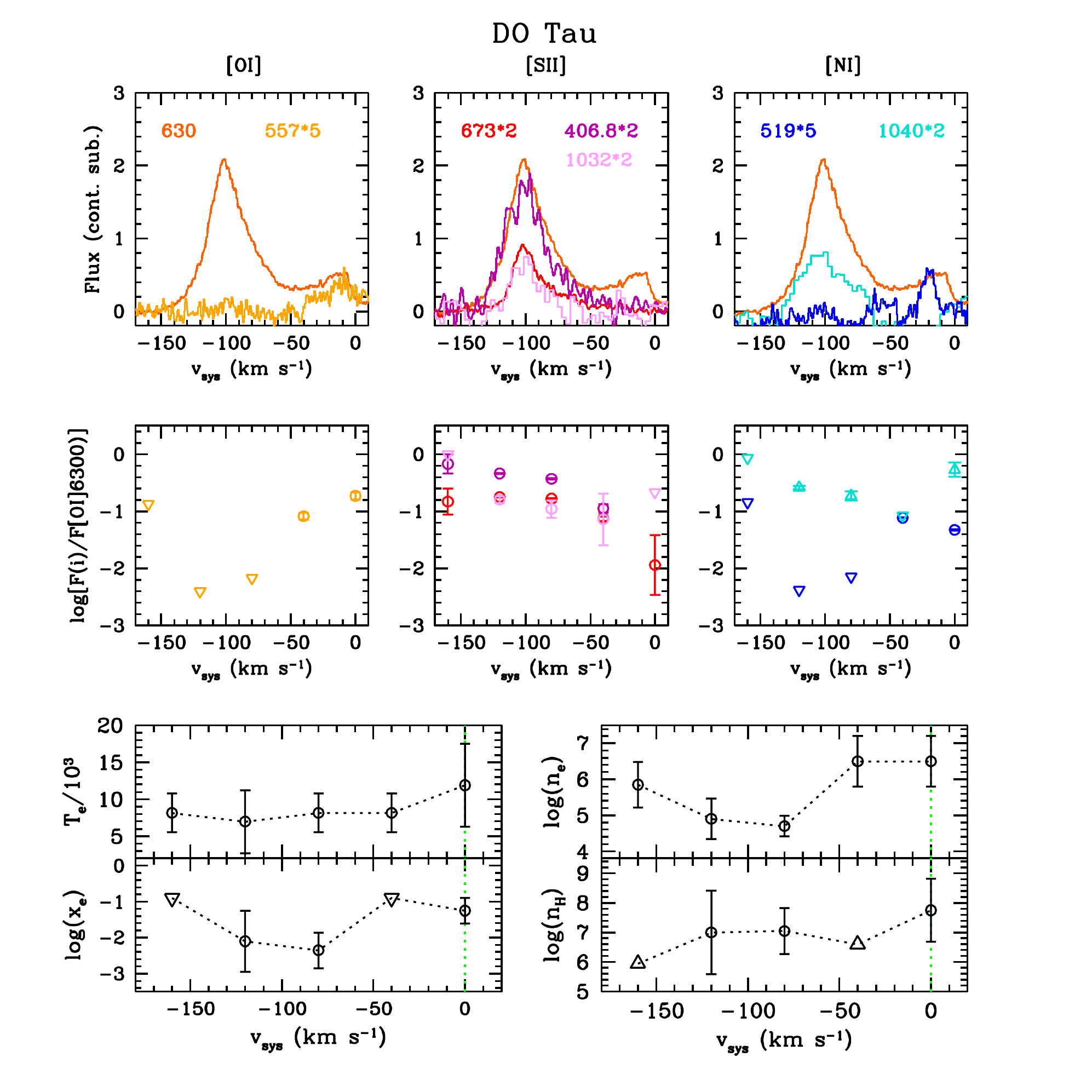}
   \caption{ \label{fig:fig12} As in Fig.\,\ref{fig:fig10} but for DO Tau. Fluxes are computed in bins of 40 km s$^{-1}$. In the middle and bottom panels, reverse (normal) triangles are 3-$\sigma$ upper (lower) limits.}
\end{figure*}

\begin{figure*}
   \centering
   \includegraphics[width=18cm]{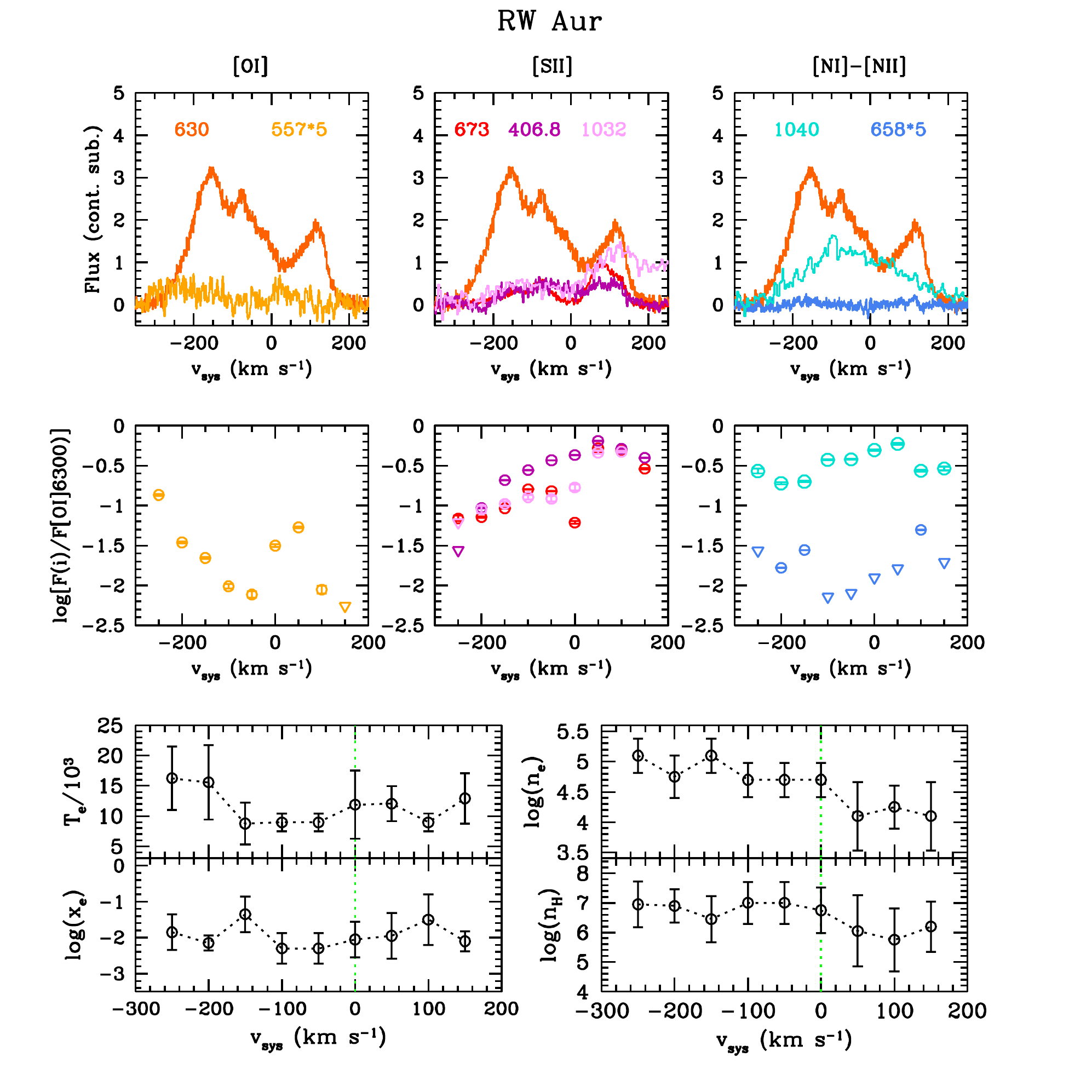}
   \caption{ \label{fig:fig13}As in Fig.\,\ref{fig:fig10} but for RW Aur. Fluxes are computed in bins of 50 km s$^{-1}$. In the middle panels, reverse triangles are 3-$\sigma$ upper limits.}
 \end{figure*}

Figures \ref{fig:fig10}-\ref{fig:fig13} show the fit results. In the upper panels, some of the observed profiles are shown in comparison with the [\oi]630 line, while in the middle
panels the de-reddened flux ratios are given as a function of velocity. The fitted \av\ is 1 mag in RW Aur, and negligible in the other sources. 
The bottom panels show the physical parameters fitted in the different velocity channels. These are also summarized in Table\,\ref{tab:tab5}.

Our main results are as follows: 1) 
In the blue-lobe of  DG Tau and  HN Tau, \Te\ increases with velocity, from about $6\,000 $ to $ 10\,000$ K at the rest velocity, up to $\sim$\,15\,000 K at the maximum  velocity.  A steep increase of \Te\, is also probed in the low-velocity, redshifted component of HN Tau.
 In RW Aur, \Te\ $\sim$\,10\,000 K at all velocities, except in the blue lobe for |$v$| \gapprox\ 200 km s$^{-1}$, where  \Te\ $\sim$ \, 15\,000 K. In DO Tau, \Te\ is roughly constant 
around 8\,000 K. 2) There is a shallow dependence of \dens\ with velocity. This is about 10$^{4.5}-$10$^5$ cm$^{-3}$ in all objects except for the LVC of DO Tau, where \dens \, $\sim$\,10$^{6.5}$ cm$^{-3}$.  3) The fractional ionization along the jet has a similar trend as the temperature. In DG Tau and HN Tau, \xe\ increases from
 several 10$^{-2}$ to 0.5$-$0.8 going from the rest velocity to $-$180 km\,s$^{-1}$. An increase of \xe\ with velocity is  also recognizable in both lobes of  RW Aur, but for values between 7 10$^{-3}$ and 4 10$^{-2}$. In DO Tau, \xe\ is \lapprox\ 5 $\times$ 10$^{-2}$ at all velocities.  4) The trend exhibited by the fractional
  ionization and the electron density influences directly  that of the total gas density \nH. This latter decreases from the rest to the apex, blueshifted velocity from about 10$^{6}$  cm$^{-3}$ to $\sim$\,10$^{5}$ cm$^{-3}$ in DG Tau and HN Tau. In RW Aur, \nH\ is $\sim$\,10$^{6}$ cm$^{-3}$ in the red lobe  and  $\sim$ 10$^7$ cm$^{-3}$ in the blue lobe. In DO Tau, \nH\ \gapprox\ 10$^{7}$ cm$^{-3}$ at all velocities.
%-------------------------------------- Figura 10
%-------------------------------------

%-------------------------------------- Figura 11
%

%-------------------------------------- Figura 12
%-------------------------------------- Figura 13

\begin{table*}
\caption{\label{tab:tab5} Physical parameters of the velocity components.}
\centering
\begin{tabular}{cccccc}
\hline\hline 
   Velocity interval   & \Te          & log(\dens)          & \xe               & log(\nH)           &  X(Fe)/X(Fe)$_\odot$     \\  
 (\kms)                & (10$^3$ K)   & (cm$^{-3}$)         &                   & (cm$^{-3}$)        &                          \\
\hline                 
{\bf DG Tau}           &              &                     &                   &                    &                          \\
  0\, $\div\,-$80      &  5.9$-$6.3   &  4.09$-$5.15        & 0.03$-$0.10       &  5.91$-$6.15       &  0.035$-$0.2             \\  
 $-$100\, $\div\,-$140 &  8.9$-$10.0  &  4.90$-$5.10        & 0.11$-$0.30       &  5.61$-$5.83       &  0.41$-$0.88             \\  
 $-$160\, $\div\,-$180 & 10.3$-$15.9  &  4.70$-$5.10        & 0.41$-$0.52       &  4.98$-$5.48       &  0.95                    \\  
 \hline
{\bf HN Tau}           &              &                     &                   &                    &                          \\
 $+$40\, $\div\,$0     & 9.8$-$14.2   &  4.40$-$4.45        & 0.04$-$0.06       &  5.65$-$5.85       &  0.06$-$0.19             \\  
 $-$20\, $\div\,-$80   & 6.7$-$9.5    &  4.70$-$4.95        & 0.06$-$0.20       &  5.65$-$5.95       &  0.21$-$0.38             \\  
 $-$100\, $\div\,-$180 & 8.0$-$13.9   &  4.95$-$5.10        & 0.41$-$0.83       &  5.03$-$5.49       &  0.65$-$1.50             \\  
 \hline
{\bf DO Tau}           &              &                     &                   &                    &                          \\
  0\, $\div\,-$40      & 8.1$-$11.9   &  6.50               & 0.055             &  7.75              &  -                       \\  
  $-$80\, $\div\,-$160 & 7.0$-$8.1    &  4.70$-$5.85        & 0.004$-$0.008     &  7.00$-$7.05       &  0.68$-$0.77             \\  
 \hline
{\bf RW Aur}           &              &                     &                   &                    &                          \\
$+$150\, $\div\,+$50   & 9.0$-$12.9   &  4.10$-$4.25        &  0.008$-$0.03     &  5.75$-$6.20       &  0.39$-$1.15             \\
  0\, $\div\,-$100     & 9.0$-$11.9   &  4.70               &  0.005$-$0.009    &  6.75$-$7.00       &  0.79$-$0.94             \\ 
$-$150\, $\div\,-$250  & 8.7$-$16.3   &  4.75$-$5.10        &  0.007$-$0.04     &  6.45$-$6.95       &  0.92$-$1.24             \\ 
\hline
\end{tabular}
\end{table*}

%\footnotesize
\begin{table*}
\caption{\label{tab:tab6} Line ratios in the zero-velocity bin compared with photoevaporative model predictions.}
\centering
\begin{tabular}{cccccc}
\hline\hline 
 Ratio                    & log(L$_X$)=30     &  DG Tau     &  HN Tau      & DO Tau     & RW Aur      \\
 \hline
 $[\oi]$557/$[\oi]$630    &  0.12        &  0.09       &  0.08        & 0.18       & 0.03        \\ 
 $[\sii]$673/$[\oi]$630   &  0.38        &  0.15       &  0.24        & 0.01       & 0.06        \\ 
 $[\sii]$407/$[\oi]$630   &  0.94        &  0.09       &  0.35        & 0.06       & 0.43        \\ 
 $[\sii]$1032/$[\oi]$630  &  0.27        &  0.06       &  0.06        & 0.23       & 0.17        \\  
 $[\n]$520/$[\oi]$630     &  0.01        &    -        &  0.03        & 0.05       & -           \\
 $[\n]$1040/$[\oi]$630    &  0.06        &  0.14       &  0.03        & 0.54       & 0.50        \\
 $[\nii]$658/$[\oi]$630   &  0.34        &  0.02       &  0.06        &   -        & 0.01        \\  
\hline                 
\end{tabular}
\end{table*}

\subsubsection{Comparison with previous results}\label{sec:sec4.1.3}

DG Tau is by far the most studied object amongst our targets. Several works have addressed the physical properties of its jet (e.g., Lavalley et al. 2000, Bacciotti et al. 2000). The most recent study by Maurri et al. (2014) in particular addressed the excitation conditions as a function of the distance from the central source and in different velocity components, namely at low, medium, and high velocity. The  applied method (Bacciotti and Eisl{\"o}ffel, 1999) involves lines not sensitive to the temperature. This latter was therefore assumed between 10\,000 K (in the LVC) and 30\,000 K (HVC). Our diagnostic analysis, however, shows that \Te\,  never exceeds 16\,000 K. The densities and ionization fractions derived by Maurri et al., averaged within about 1$\arcsec$, which is roughly our resolution, are consistent with our velocity-averaged values reported in Table\,\ref{tab:tab5}.  In particular, their electron density and ionization fraction are low in the LVC and increase up to \dens \,$\sim\,$5\, $\times$ $10^4$ cm$^{-3}$ and \xe\,\,$\sim$\,0.5 in the HVC.  
It is however to be noted that the kinematics of the DG Tau jet has significantly changed from the observations reported by Maurri et al. (taken in January 1999) and that their HVC and MVC do not refer to the same velocity ranges considered by us. In particular, their HVC comprises gas between $-$400 and $-$300 \kms , while in our observations the HVC reaches about $-$200 \kms\, at maximum. The fact that the physical conditions are preserved even if the kinematics has changed suggests that the jet has undergone a deceleration over time, although maintaining the same conditions of excitation. In the context of shock excitation, this implies that the shock velocity has remained fairly constant. One possibility is that during the last 20 years the pre-shock density has increased due to a piling-up of previous bow-shocks with time (Raga et al. 1998), and that in the impact with the ambient medium the jet energy is now mainly transformed into gas heating, with a consequent decrease in speed.

The excitation conditions of the RW Aur jet have been studied in  Melnikov et al. (2009). They separately analyzed the blue- and redshifted components of the jet, finding that the redshifted component is denser (when considering the total density) and less ionized than the blueshifted component. We do not see this significant difference in the ionization fraction that remains very low ($<$0.1) in both lobes. Our derived electron density is higher than that measured by Melnikov et al. by up to an order of magnitude,
likely because their determination relies on the [\sii]673/671 ratio, which is sensitive only to low-density regimes.

The physical parameters of the low-velocity component, associated with slow winds, were recently analyzed in T Tauri stars (e.g., Natta et al. 2014, Fang et al. 2018). Both studies suggest that gas in the LVC is characterized by very high electron densities (i.e., \dens\,$\sim$\,10$^7\,-$\,10$^8$ cm$^{-3}$) and moderate temperatures (i.e., 5000\,$-$\,10\,000 K). Their analysis is based on the two line ratios [\oi]630/557 and [\sii]406.9/[\oi]630. The latter ratio is however relatively insensitive to variations of density and temperature (see Fig.\,\ref{fig:fig9}) so it is in practice difficult to remove the degeneracy of the physical parameters using only these ratios. Through our multi-line analysis,  for
the LVC we generally derive a temperature between 6000 and 10\,000 K, an electron density  $\sim$\,10$^4$\,$-$\,10$^5$ cm$^{-3}$ (except for DO Tau), and an ionization fraction less than 0.1. Correspondingly, the total density of the LVC is very high (\nH\,$\sim$\,10$^6\,-$\,10$^7$ cm$^{-3}$), that is, always higher than in the components at larger velocities.

\subsection{Iron depletion}\label{sec:sec4.2}
An important observational constraint for models of jet and wind formation is the estimate of the amount of dust inside the jet beam.  
 Dust-free jets are predicted by models if the launching zone is in the gaseous disk close to the star where the dust is destroyed by stellar radiation. Vice versa, a significant amount of dust is expected in winds that originate in disk regions beyond the sublimation radius, or in shocks inefficient in destroying the dust through vaporization or sputtering processes (Jones 2000, Guillet et al. 2009). The partial or total disruption of dust implies the release of species in the gaseous phase, including iron, locked in grain material in quiescent conditions. Therefore, the abundance of iron in gas-phase (X(Fe)) is an indirect measure of the content of dust inside the flow.

\begin{figure*}
   \centering
   \includegraphics[width=16cm]{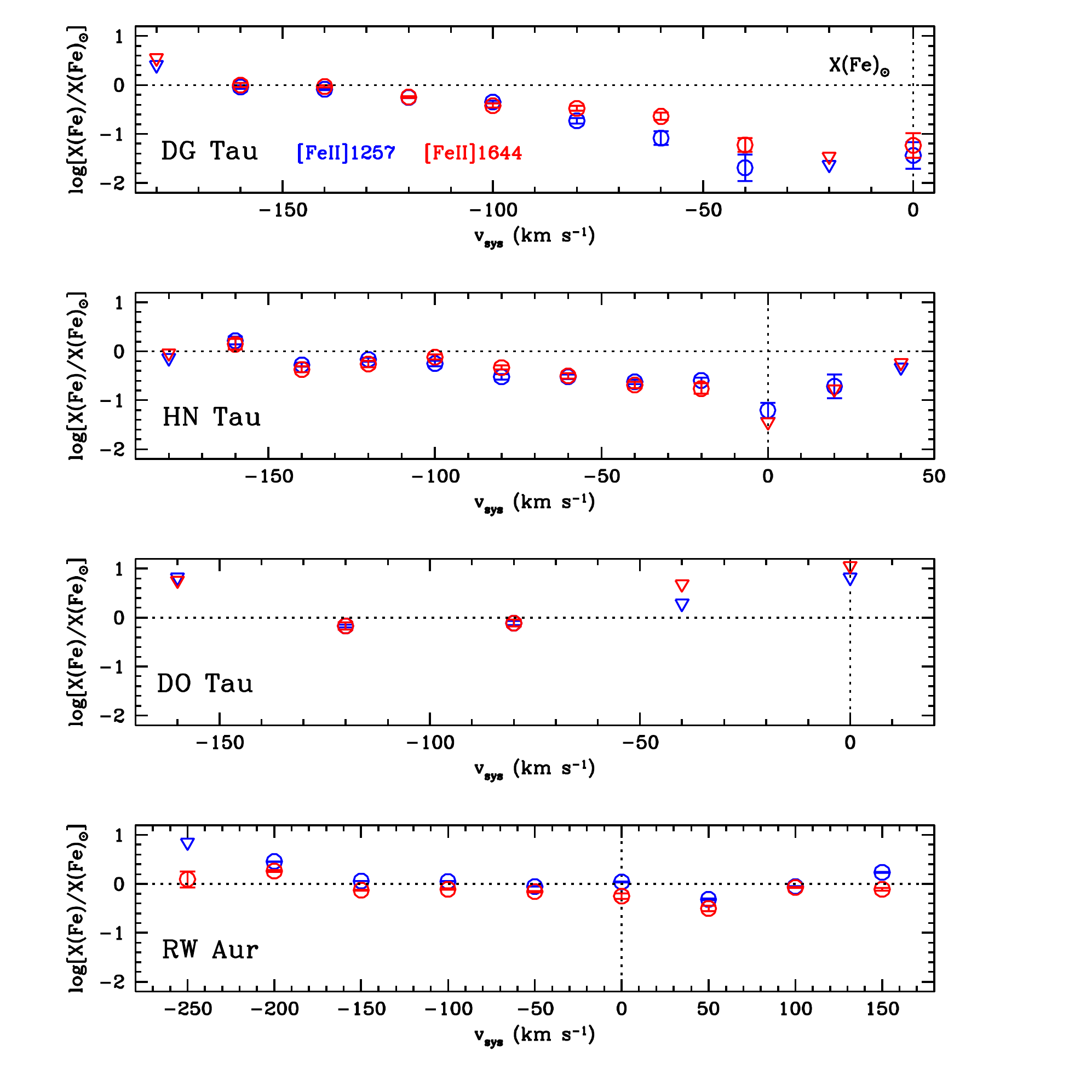}
   \caption{ \label{fig:fig14} From top to bottom: log[(X(Fe)/X(Fe)$_{\odot}$)] as a function of velocity for  DG Tau, HN Tau, DO Tau, and RW Aur.
   Blue and red points are the values estimated from the [\feii]1257 and [\feii]1644 line, respectively.  Reverse triangles are 3$\sigma$ upper limits. In each panel, the horizontal dotted line indicates the solar abundance value. The vertical line marks the rest velocity.}
 \end{figure*}

Several methods based on different line ratios have been proposed to measure X(Fe) in nebular regions: [\feii]1257/[\pii]1189 (Oliva et al. 2001, Nisini et al. 2005, Giannini
 et al. 2008, Podio et al. 2006),  [\feii]1257/Pa$\beta$ (Nisini et al. 2002), and  [\feii] UV lines over [\oi]630 (Giannini et al. 2013). In this work we
  adopt the [\feii]1257,1644/[\oi]630 ratios, with the assumption of solar oxygen abundance (Asplund et al. 2005). We have applied our excitation and ionization model of \feii\, to predict the [\feii]1257 and [\feii]1644 line emissivity assuming the physical conditions derived in the various velocity channels from the diagnostic analysis. The theoretical ratios are compared with the observed fluxes to get a measure of the percentage of iron in the gas phase, namely X(Fe)/X(Fe)$_{\odot}$. The results are shown in Fig.\,\ref{fig:fig14} and summarized in the last column of Table\,\ref{tab:tab5}. 
A remarkable depletion of iron is present at low velocities in DG Tau and HN Tau. The minimum value is found at $v$ = $-$40 km\,s$^{-1}$ in DG Tau, where X(Fe)$\sim$3.5\% X(Fe)$_{\odot}$
and at zero velocity in HN Tau, where X(Fe)\,$\sim$\,6\% X(Fe)$_{\odot}$. In both objects, X(Fe) increases with velocity, and reaches values compatible with
 the solar ones in the HVC. We measure some iron depletion also at low velocity in the red lobe of RW Aur (X(Fe)$\sim$40\% X(Fe)$_{\odot}$), while in all 
 the other velocity bins we get X(Fe) \lapprox  X(Fe)$_{\odot}$. Finally, in DO Tau we can derive X(Fe) only in two velocity bins centered at $-
 $80 km\,s$^{-1}$ and $-$120 km\,s$^{-1}$, where X(Fe)$\sim$ 70\% X(Fe)$_{\odot}$. At low velocity we can only estimate upper limits, which are however not significant.

%-------------------------------------- Figura 14

\section{The emerging picture} \label{sec:sec5}
The picture that emerges from our diagnostic analysis is that of a flow in which  the physical parameters vary smoothly with velocity. This is particularly evident in DG Tau and HN Tau, characterized by high temperature and ionization at high velocity, which gradually diminish until the gas becomes mainly neutral at low velocity. Also, the total density decreases with velocity, from about  $5\,\times$ 10$^5-$10$^6$ cm$^{-3}$ at the rest velocity to a few 10$^5$ cm$^{-3}$ at the highest blueshifted speed. 
In particular, a high ionization fraction (up to about 0.6$-$0.8) is attained in the high velocity range. As already discussed in previous studies, 
such a high ionization fraction  is more consistent with shock heating compared to other mechanisms; for example, ambipolar diffusion, which is unable to induce a significant ionization (Garcia et al. 2001). 
So far, the model that has proven to be the most consistent with various observables is the so-called disk-wind model (e.g., Ferreira et al. 2006), in which the jet is launched from a region involving a large range of disk radii. In DG Tau, for example,
Coffey et al. (2007) estimated that the HVC  is launched from a distance of 0.2$-$0.5 au from the star along the disk plane, while the LVC from a distance as far as 1.9 au. Further support to MHD models comes from high-angular-resolution observations, which show that the jet width associated with the low-velocity gas component is larger than the high-velocity one (Maurri et al. 2014, Agra-Amboage et al. 2011). 
The MHD scenario naturally explains  the smooth variation of parameters as a consequence of a gradient in the physical conditions across the jet width: the inner, high-velocity, jet streamlines are more ionized than the outer, almost neutral regions at low velocity, while the total density  is higher in the outer regions of the jet. This is also predicted by some MHD models, such as the Romanova et al. (2009) jet model, where a high-density conical low-velocity wind is predicted to surround a low-density and high-velocity axial jet. In this scenario, the remarkable iron depletion observed at low velocity can be explained either with a radial velocity gradient across the disk, with higher and more ionized gas located inside the sublimation radius, or as a consequence of the gas acceleration, as higher-velocity shocks are more efficient in destroying the dust grains.

DO Tau and RW Aur show remarkable differences with respect to the other two sources, as for them the gas ionization never rises above \xe\,=\,0.1. While this value agrees better with the predictions of models for ambipolar diffusion, a strong discrepancy remains between the predicted electron densities ($\sim$ 10$^3$ cm$^{-3}$, Garcia et al. 2001) and our estimates ($\sim$ 10$^5$ cm$^{-3}$).

In DO Tau, the low ionization fraction may be explained with a low-velocity shock.  When corrected for the inclination angle (Table\,\ref{tab:tab1}), the maximum velocity of the DO Tau jet is only 180 km\,s$^{-1}$ (being between 250 and 580 km\,s$^{-1}$ in the other objects), and the width of the lines in the HVC are small (Table\,\ref{tab:tab4}), as expected for a low-velocity shock (Hartigan et al. 1987). Also, the temperature remains low at all velocities. 
Conversely, in RW Aur, although the ionization is low, the temperature exceeds 10\,000 K  and the line wings extend up to 250 km\,s$^{-1}$. Noticeably, DO Tau  and RW Aur are the objects with the highest total density, namely up to  10$^7$ cm$^{-3}$. Indeed, different pre-shock densities can explain the difference in ionization fraction we observe in our sources (e.g., Hartigan et al. 1994). 
One remaining question pertains to the DO Tau and RW Aur jets being significantly denser than those of DG Tau and HN Tau. In our sample, the stellar parameters (Table\,\ref{tab:tab1}) are not significantly different from each other, and do not seem to be correlated with the jet densities or other excitation parameters.
%Mass accretion rates span about one order of magnitude in the objects of our sample, but does not seem to be correlated with the jet densities or other excitation parameters. 
The only difference we note between DO Tau, RW Aur, and the other sources is in the jet opening angles in the inner region. In fact, 
 the width of the DG Tau jet at a  distance of 50 au from the source varies between 15 and 35 au, depending on the velocity range (Maurri et al. 2014), and it is similar to the velocity-integrated width in HN Tau (20 au; Hartigan et al. 2004). Conversely, the velocity-integrated jet widths of both RW Aur and DO Tau at 50 au are less than 10 au (Woitas et al. 2002, Erkal 2018). On this basis, one possibility is that the lower density we observe in the DG Tau and HN Tau jets with respect to DO Tau and RW Aur is caused by a quicker drop in density as the jets propagate and expand in their collimation cone. \\
 Alternatively, in DO Tau and RW Aur, the jets could originate in dense and almost neutral regions of the gaseous disk. This could be the case in particular for RW Aur, whose distance of the jet footprint has been estimated to be within 0.5 and 1.6 au from the star for the blue and red lobes, respectively (Woitas et al. 2005). 
Albeit only qualitatively, this scenario could also explain why no iron depletion is observed in these sources.

Finally, we note that the smooth variation of the physical parameters makes a different origin for the LVC and HVC unlikely. This is predicted, for example, by disk photoevaporative models  (e.g., Ercolano and Owen 2016 and references therein) that interpret the LVC  as a slow wind originating on the disk surface due to the irradiation of the stellar UV/X photons. These models predict bright forbidden lines collisionally excited in the outflowing wind. For example, in Table\,\ref{tab:tab6} we show the comparison of some observed ratios in the zero-velocity bin with the predictions for a photoevaporative disk-wind irradiated by stellar X-rays (log\rm{(L$_X$/\lsun\,)}\,=\,30  Ercolano, priv. comm.). It is interesting that while there is a fair agreement for line ratios of neutral species ([\oi] and [\n] lines), a remarkable discrepancy is found when considering ratios involving ionized species ([\nii] in particular). This indicates that the theoretical gas ionization is significantly higher than observed in the low-velocity component of our jets. With reference to the analysis of Appendix A, it is relatively likely that the stellar irradiation is only able to ionize the species with very low ionization potentials (e.g., S and Fe). Noticeably, the reported results do not change significantly, even decreasing the stellar X-ray luminosity by two orders of magnitude (Ercolano and Owen 2010, their Table\,1).

\section{Summary}\label{sec:sec6}
In the framework of the GHOsT project, we present GIARPS observations of a sample of six jets from T Tauri stars. The high spectral resolution coupled with the wide wavelength range coverage allowed us to trace the variations of the gas parameters with velocity. In four objects,  DG Tau, HN Tau, DO Tau, and RW Aur, we  observed many atomic forbidden lines
of [\oi]\,, [\sii]\,, [\n]\,, [\nii]\,, and [\feii]. Our results can be summarized as follows:
\begin{itemize}
\item[-] In all objects, we detect the blueshifted line components, which typically extend up to $-$200  km\,s$^{-1}$.  Redshifted gas is seen only at low
velocities (\lapprox + 40 km\,s$^{-1}$), with the exception of RW Aur, in which the redshifted component is detected up to $+$150 km\,s$^{-1}$. Apart from a few cases, the
 line profiles have a complex shape, typically presenting a blueshifted LVC, and an HVC that peaks at more than $-$100 km\,s$^{-1}$. The LVC is
  preferentially bright in low-excitation lines of neutral species, while a HVC is bright also in lines of ionized species. 
\item[-] Line ratios of DG Tau, HN Tau, DO Tau, and RW Aur were analyzed through an NLTE excitation model combined with an ionization model. From this, we are able to infer the variation of the excitation conditions  of the gas (temperature, fractional ionization, electron, and total density) with the jet acceleration. 
\item[-] All the physical parameters smoothly change with velocity, therefore suggesting a common emission mechanism for the LVC and the HVC.
\item[-] DG Tau and HN Tau share similar excitation conditions. In both objects, temperature and fractional ionization rise with velocity, with \Te\, $\sim$ 6000$-$8000 K and \xe\, \lapprox\, 0.1 in the LVC and \Te\, $\sim$ 15\,000 K and \xe\, $\sim$ 0.6-0.8 at the maximum blueshifted velocity. In both these objects the electron density is $\sim$ 10$^{4.5-5}$ cm$^{-3}$, and the total density  $\sim$ 10$^5-$10$^6$ cm$^{-3}$, with a peak in the LVC.
\item[-]Unlike the case for DG Tau and HN Tau, the fractional ionization in DO Tau and RW Aur is very low (10$^{-2}-$10$^{-3}$), and the temperature is almost constant at all velocities ($\sim$ 8000 K in DO Tau and $\sim$ 10\,000 K in RW Aur). The total density is remarkably high, ranging between  $\sim$ 10$^6$ cm$^{-3}$ and more than 10$^7$ cm$^{-3}$.
\item[-] From an estimate of the iron abundance in the gaseous phase we probed the dust content in the jets. In DG Tau and HN Tau,  X(Fe) increases with velocity, going from less than 10\% X(Fe)$_\odot$ in the low-velocity bins to  $\sim$ X(Fe)$_\odot$ at the apex velocity. Although not so evident, some signs of iron depletion are also recognizable in the redshifted, low-velocity component of RW Aur, while in DO Tau, X(Fe) \lapprox X(Fe)$_{\odot}$ for $v$ $\approx$\,$-$100 km\,s$^{-1}$, but this is not constrained at lower velocity.
\end{itemize}

In conclusion, despite the fact that the investigated sample is composed of only six objects, the physical conditions we derive differ significantly from each other. DG Tau and HN Tau  are the only two objects that share similar physical conditions. The temperature and ionization gradients observed in these objects favor MHD shock heating, in which the warmer and more ionized streamlines originate in the internal and mainly gaseous disk, while the low-velocity and almost neutral streamlines come from the dusty regions of the outer disk. The other two sources, and DO Tau in particular, are characterized by a very low ionization degree, a high total density, and an almost negligible iron depletion degree.  These results have been tentatively explained by the formation of these jets from dense regions inside the gaseous inner disk, or alternatively as a consequence of a high degree of collimation. Why such differences exist among our objects remains an open question, to be investigated on a more robust statistical basis. Finally, stellar irradiation, although responsible for the ionization of species like S and Fe, is discarded as the excitation mechanism of the gas in the disk wind.

\begin{appendix}
\section{The excitation and ionization model}\label{sec:sec4.1.1}
The excitation model assumes a NLTE approximation for line emission and the  population of levels is determined by assuming equilibrium between collisional excitation and de-excitation with electrons, and radiative decay. We developed a five-level model for O$^0$, S$^+$, N$^0$, and N$^+$, whose details about the adopted radiative and collisional rates are described in Giannini et al. (2015). The Fe$^+$ model includes 159 levels whose radiative and collisional rates  are taken from Bautista and Pradhan (1998).

\begin{figure*}
   \centering
   \includegraphics[width=14cm]{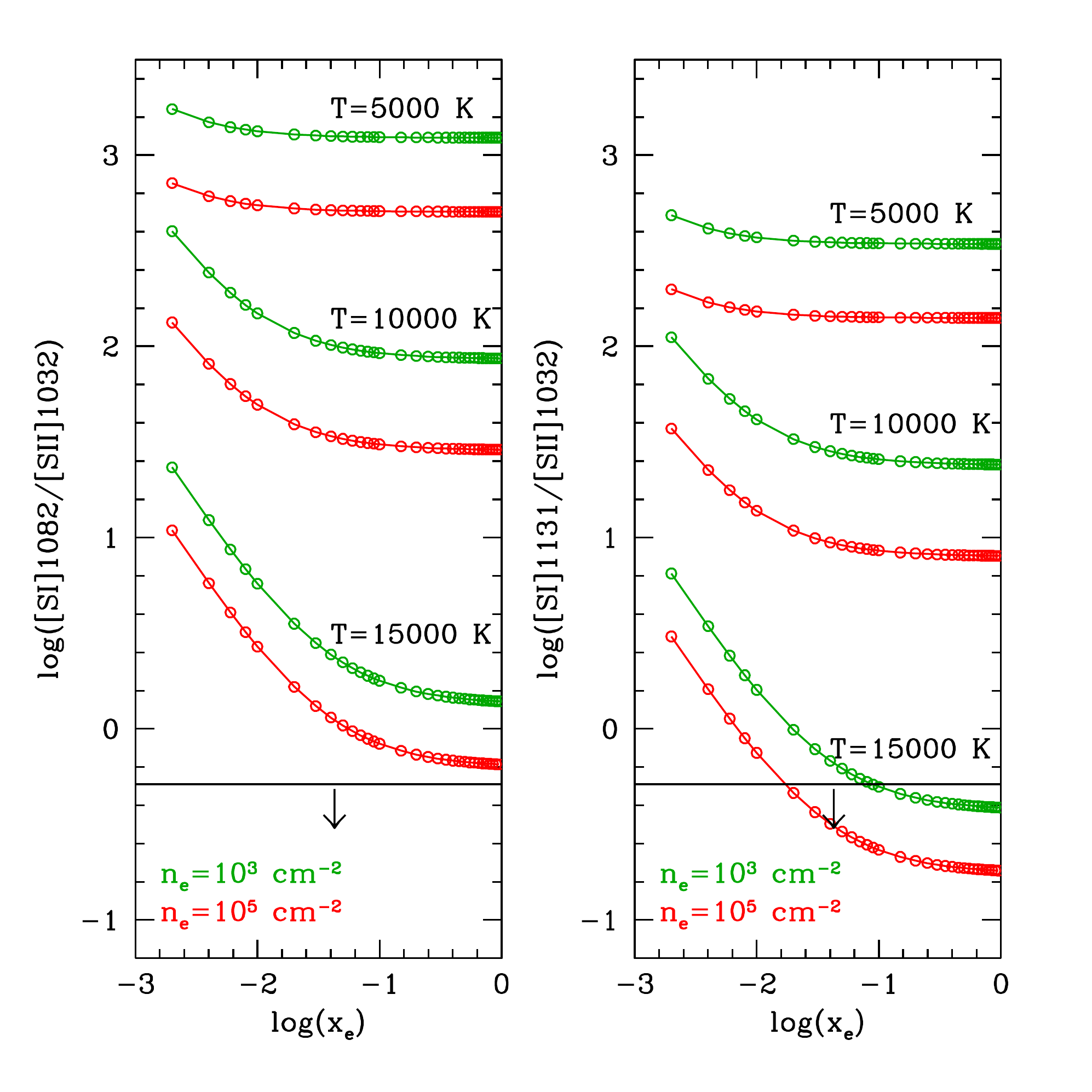}
\caption{\label{fig:fig8} Theoretical  [\si]/[\sii] flux ratios vs. \xe\, obtained assuming  S= \si +\sii+\siii. Noticeably, both the [\si]
lines are predicted to be brighter than the [\sii]1032 line in the range of \Te\, and \dens\, that we fit (see Sect.\ref{sec:sec4}). The horizontal line is the typical 3-$\sigma$ upper limit measured on our objects. }
 \end{figure*}

The ionization model includes the following processes:
collisional ionization, radiative and dielectronic recombination, and direct and inverse charge-exchange with hydrogen (see Giannini et al. 2015 for details).  We consider the ionization equilibrium equations for the first three ionic stages of each atomic species (O,\,S,\,N,\,Fe). 
There is however evidence that no neutral sulfur is present in the gas from which the outflow originates, since no [\si] lines are detected in our spectra. In Fig.\,\ref{fig:fig8}
we plot the ratio between [\si]1082.4/1130.8 and [\sii]1032 against \xe, as computed with our model for different temperatures and densities. Noticeably, both [\si]\, lines are predicted up to three orders of magnitude brighter than [\sii]1032 and the 3-$\sigma$ upper limits estimated in our spectra are consistent with the theoretical ratios only in the unrealistic case of a gas with high temperature but very low ionization degree. The most plausible explanation is that the jet and winds we observe are launched from the more external layers of the inner disk, where species of low ionization
potential (IP(S)=10.36 eV) are easily ionized by the stellar FUV photons (e.g., Gorti and Hollenbach 2008). Indeed, the only [\si] line ever detected in young stars is the [\si] fundamental line at 25.25 $\mu$m, observed by Spitzer in embedded sources shielded against the stellar FUV radiation (e.g., Dionatos et al. 2009). This line, however, has never been detected in T Tauri stars (e.g., Lahuis et al. 2007), where plausibly the shielding effect is less efficient.\\ 
Similarly, we expect that iron is also mostly or fully ionized, because its ionization potential is even lower than that of sulfur (IP(Fe)=7.90 eV).

\begin{figure*}
   \centering
\includegraphics[width=14cm]{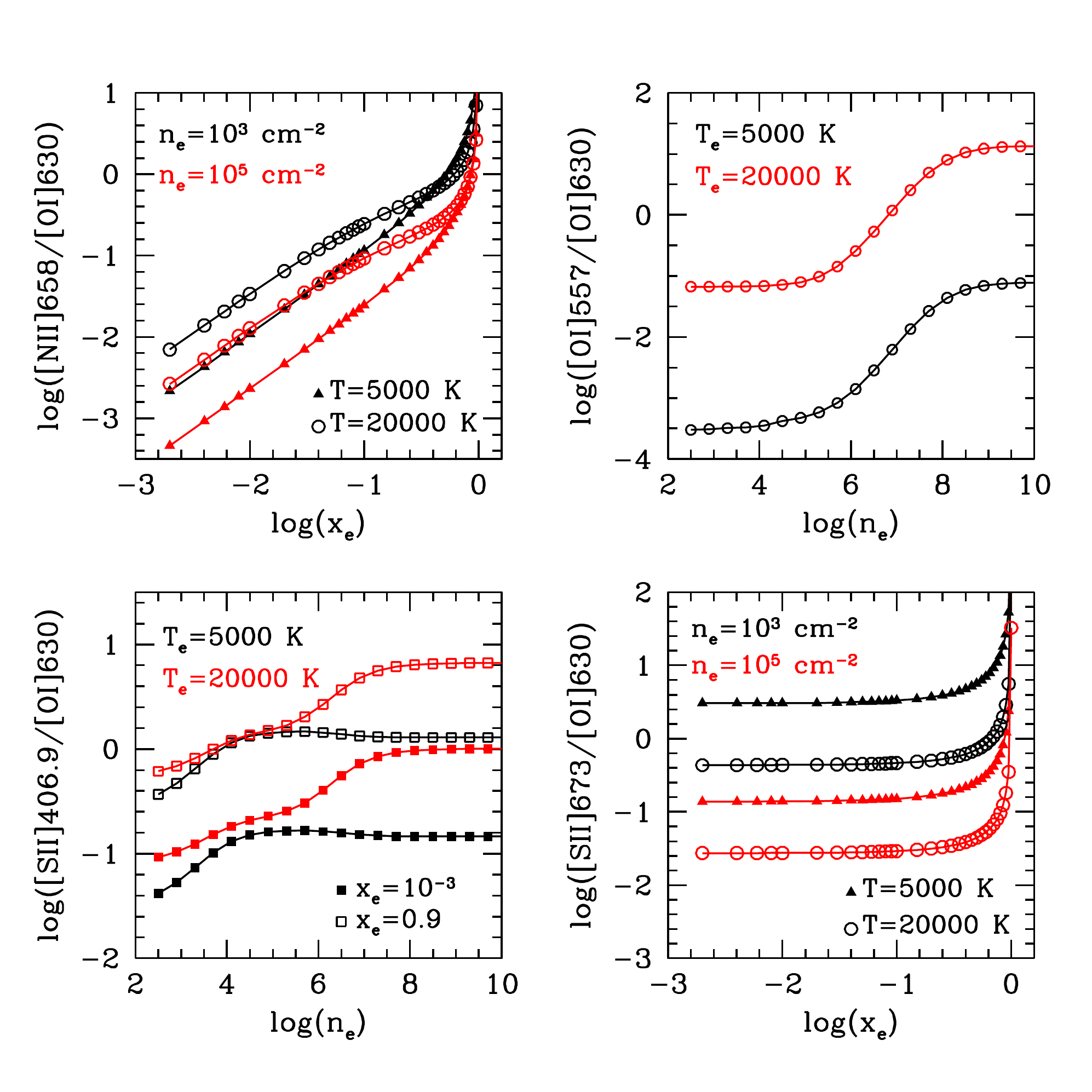}
\caption{\label{fig:fig9} Diagnostic diagrams of line ratios sensitive to the physical parameters. Solar abundance is assumed for all the involved species. }
\end{figure*}

As output from the excitation and ionization model we get theoretical line emissivities as a function of the  electron temperature and density (\Te\,, \dens), and the fractional ionization \xe\,=\,$n_{\rm{e}}/n_{\rm{H}}$,  where $n_{\rm{H}}$\,=\,$n_{\rm{H^0}}$+$n_{\rm{H^+}}$. 
As an example, we show in Fig.\,\ref{fig:fig9} some of the flux ratios that are significantly sensitive to the gas parameters. The fractional ionization is well constrained by the 
[\nii]658/[\oi]630 ratio (upper left panel), that spans over about three orders of magnitude for 10$^{-3}\le$\,\xe\,$\le$\,1, with a shallow dependence on \Te\, and \dens.
The [\oi]557/[\oi]630 ratio (upper right panel) is a good probe of the electron density in the range 10$^4$ cm$^{-3}$\,$\le$\,\dens\,$\le$\,10$^8$ cm$^{-3}$, but it is also depends on \Te\,. Other ratios, such as [\sii]408.9/[\oi]630 (bottom.left panel) even depend on all the three parameters. 

We run the excitation and ionization model to compute a grid of theoretical line emissivities 
 for the [\oi], [\sii], [\n], and [\nii] lines of Table\,\ref{tab:tab3}. Solar abundances have been assumed (Asplund et al. 2005). The parameter space is : 
 4000 K $\le$ \Te\, $\le$ 50000 K (in steps of 2000 K for \Te\, $\le$\, 20000 K and 5000 K for \Te\, $>$ 20000 K), 10$^2$ cm$^{-3}$\,$\le$ \dens\, $\le$\, 10$^{10}$ cm$^{-3}$ (in steps of log$_{10}$ ($\delta$\dens\,/cm$^{-3}$) = 0.1), and 10$^{-3}$ $\le$\, \xe\,$\le$ 1 (in steps of 10$^{-3}$, 10$^{-2}$ and 10$^{-1}$, for 10$^{-3}$ $\le$\, \xe\, $\le$ 10$^{-2}$,   10$^{-2}$ $\le$\, \xe\, $\le$\, 10$^{-1}$, 0.1 $\le$\, \xe\, $\le$ 1, respectively).

%-------------------------------------- Figura 8
%q

%-------------------------------------- Figura 9

\end{appendix}

\begin{acknowledgements}
The authors are very grateful to Barbara Ercolano for providing the predictions of her photoevaporative model.
 We thank S. Dallaporta (ANS Collaboration) for his BVRI observations of our targets and C. Manara for his comments.
This work has been supported by the project PRIN-INAF 2016
The Cradle of Life - GENESIS-SKA (General Conditions in
Early Planetary Systems for the rise of life with SKA).
SA acknowledges the support by INAF/Frontiera through the "Progetti Premiali" funding scheme of the
Italian Ministry of Education, University, and Research.
\end{acknowledgements}

\end{document}